\documentclass[fleqn,usenatbib]{mnras}

\usepackage{newtxtext,newtxmath}

\usepackage[T1]{fontenc}

\DeclareRobustCommand{\VAN}[3]{#2}
\let\VANthebibliography\thebibliography
\def\thebibliography{\DeclareRobustCommand{\VAN}[3]{##3}\VANthebibliography}

\usepackage{graphicx}	
\usepackage{amsmath}	

\title[ARTEMIS emulator]{ARTEMIS emulator: exploring the effect of cosmology and galaxy formation physics on Milky Way-mass haloes and their satellites}

\author[S. T. Brown et al.]{
Shaun T. Brown,$^{1}$\thanks{E-mail: shaun.t.brown@durham.ac.uk}
Azadeh Fattahi,$^{1}$ Ian G. McCarthy,$^{2}$ Andreea S. Font,$^{2}$ Kyle A. Oman$^{1,3}$ and \newauthor Alexander H. Riley$^{1}$
\\
$^{1}$Institute for Computational Cosmology, Department of Physics, Durham University, Durham DH1 3LE, UK\\
$^{2}$Astrophysics Research Institute, Liverpool John Moores University, 146 Brownlow Hill, Liverpool L53RF, UK\\
$^{3}$Centre for Extragalactic Astronomy, Department of Physics, Durham University, Durham DH1 3LE, UK\\
}

\date{Accepted XXX. Received YYY; in original form ZZZ}

\pubyear{2024}

\begin{document}
\label{firstpage}
\pagerange{\pageref{firstpage}--\pageref{lastpage}}
\maketitle

\begin{abstract}
We present the new ARTEMIS Emulator suite of high resolution (baryon mass of $2.23 \times 10^{4}$ $h^{-1}$M$_{\odot}$) zoom-in simulations of Milky Way mass systems. Here, three haloes from the original ARTEMIS sample have been rerun multiple times, systematically varying parameters for the stellar feedback model, the density threshold for star formation, the reionisation redshift and the assumed warm dark matter (WDM) particle mass (assuming a thermal relic). From these simulations emulators are trained for a wide range of statistics that allow for fast predictions at combinations of parameters not originally sampled, running in $\sim 1$ms (a factor of $\sim 10^{11}$ faster than the simulations). In this paper we explore the dependence of the central haloes’ stellar mass on the varied parameters, finding the stellar feedback parameters to be the most important. When constraining the parameters to match the present-day stellar mass halo mass relation inferred from abundance matching we find that there is a strong degeneracy in the stellar feedback parameters, corresponding to a freedom in formation time of the stellar component for a fixed halo assembly history. We additionally explore the dependence of the satellite stellar mass function, where it is found that variations in stellar feedback, the reionisation redshift and the WDM mass all have a significant effect. The presented emulators are a powerful tool which allows for fundamentally new ways of analysing and interpreting cosmological hydrodynamic simulations. Crucially, allowing their free (subgrid) parameters to be varied and marginalised, leading to more robust constraints and predictions.

\end{abstract}




\section{Introduction} \label{section:intro}

Cosmological hydrodynamic simulations have become an invaluable tool to model the formation and evolution of galaxies across a wide range of spatial and temporal scales. These simulations are able to follow the non-linear evolution of matter from the very early universe through to today, self-consistently modelling the effects of gravity, hydrodynamics and key astrophysical processes, such as star formation and feedback, in a fully cosmological context (see \citealt{Vogelsberger_20} for a recent review of the key ingredients in modern cosmological galaxy formation simulations). While early simulations were in poor agreement with observations, producing galaxies that were too massive, too compact and formed too early \citep[e.g.][]{Scannapieco_12}, it is now routine for many simulations to create realistic populations of galaxies over a wide range of masses and redshifts that match a diverse range of observed scaling relations. A non-exhaustive list includes EAGLE \citep{Eagle,Eagle_cal}, Illustris(-TNG) \citep{Illustrious,Ilustrious-tng}, Simba \citep{Simba}, FIRE(-Box) \citep{Fire-2,Firebox}, Horizon-AGN \citep{Horizon_agn} and Romulus \citep{Romulus}.

While current simulations have made great progress over the past few decades, these successes are not derived from first principles. Instead, due to the limited resolution of these types of simulations, many key processes, such as stellar and AGN feedback, are implemented through numerical routines that aim to effectively mimic the impact of these physical processes. These `subgrid' routines introduce a number of free parameters, with some having clear physical analogues, and can therefore be constrained by current observations, while others are numerical in nature with no clear observable analogue. It is common to constrain these parameters such that the simulated galaxy population matches a range of chosen observables, a process often referred to as calibrating the simulations. Thus, the success of a particular simulation is dependent on both the model itself, as well as the calibration approach. Due to the high computational expense of these simulations, calibration is often performed by running a relatively small number of development simulations used to explore the available parameter space, then choosing a combination of parameters that gives a desired fit to a set of observables. One limitation of this approach is that it is often unclear if the chosen combination of parameters is optimal, or if there are strong degeneracies within the parameter space, in turn limiting the predictive power of the simulations.

While it is necessary to consider the uncertainties, and potential freedoms, in the subgrid parameterisation when studying their effect on galaxy formation and evolution, it is equally important to consider when using such simulations to constrain different cosmological models. This is particularly relevant at small scales, where there have been tensions between the predictions of simulations that assume the standard cold dark matter (CDM) model and observations of the local universe, such as the cusp-core problem \citep[e.g.][]{flored,Moore_94}, the missing satellites problem \citep[e.g.][]{Klypin_99,Moore_99} and the too big to fail problem \citep[e.g.][]{Boylan-Kolchin_11} (see \citealt{Bullock_boylan} for a review). However, it is now well established that the inclusion of baryonic processes, such as supernova, stellar winds and AGN feedback and reionisation, plays a significant role on small scales and is able to alleviate, and potentially resolve, these tensions within the standard $\Lambda$CDM cosmological model \citep[e.g.][]{Sales_22}. However, many of these conclusions are based on using subgrid models and parameters that have been developed, and calibrated, assuming CDM. Therefore, while such conclusions suggest CDM is one potential explanation of the observations, it does not sufficiently show that CDM is a unique solution, where it is possible that alternative cosmological models, such as warm dark matter (WDM) \citep[e.g.][]{Lovell_14}, self interacting dark matter \citep[e.g.][]{Kaplinghat_16} or fuzzy dark matter \citep[e.g.][]{axion}, may also be able to describe the observed data, but with different choices of baryonic (subgrid) parameters.

The key factor limiting a full exploration of the available parameters space, and the use of more statistically rigorous techniques to do this, is the large computational expense of these types of simulations (typically $\sim 10^3$--$10^6$ cpu-hours). A promising alternative is to instead develop emulators that allow for fast predictions without having to directly run a simulation. Within large scale structure (LSS) cosmological analysis the use of such techniques is becoming commonplace. Here, emulators have been developed to reproduce the cosmological dependence predicted from N-body simulations for a range of LSS statistics, such as the non-linear matter power spectrum \citep[e.g.][]{Coyote_pk,Upadhye_14, Mira-titan, Giblin_2010}, or the halo mass function and galaxy clustering \citep[e.g.][]{Dark_quest, Bacco}. There are also a number of works that have used emulation to explore the effect of variations to the assumed galaxy formation parameters. As examples, \citet{Bower_10} use emulation in the context of the GALFORM semi-analytic model to explore the effect a range of galaxy formation parameters have on the predicted luminosity functions, and both the FLAMINGO \citep{Flamingo,Flamingo_cal} and Romulus \citep{Romulus} hydrodynamic simulations use emulators (or very similar methods) to calibrate their galaxy formation (subgrid-)parameters. So far, few works have studied the joint effect of varying the cosmological and baryonic (subgrid-) parameters, with a notable exception being the CAMELS simulations \citep{Camels} that vary some of the Friedmann parameters alongside feedback (subgrid) parameters within the Illustris-TNG model.

In this paper we present a new suite of simulations developed to explore joint variations in both the baryonic (subgrid) implementation and the assumed cosmological model. We present a suite of high resolution ($\sim 10^4 M_{\odot}$ in particle mass) Milky Way-mass zoom-in simulations, where a number of haloes (originally from the ARTEMIS sample; \citealt{Font_2020,Font_2021,Font_2022}) have been resimulated many times, systematically varying the WDM mass alongside the stellar feedback parameters, the star formation threshold and the assumed redshift of reionisation. These parameters have specifically been chosen as they all have a notable effect on the formation and evolution of the properties of the satellites to the Milky Way (i.e. dwarf galaxies). From the simulations we construct machine learning emulators that allow for fast ($\sim 1$ms) predictions of a diverse range of statistics for combinations of parameters that were not sampled originally. The significant increase in computation speed, a factor of $\sim 10^{11}$, fundamentally changes the type of analysis that is possible, allowing a full exploration of the available parameter space and marginalising over the baryonic (subgrid) parameters when making cosmological constraints and significantly improving the robustness and predictive power of the simulations.

In this first paper we present the new simulation suite and the emulators, alongside our initial results and analysis. In Section~\ref{section:sim_details} we describe the technical details of the simulations, focusing on the physical parameters of the model that are varied. In Section~\ref{section:emulator} we describe how the parameters are systematically varied and sampled with simulations, in total presenting $97$ simulations that are used for training and evaluation. We then describe how these simulations are used to build emulators using Gaussian processes for a wide range of statistics, for both the host and satellite populations, evaluating their performance. In Section~\ref{section:analysis} we explore how the stellar mass of the host galaxies (i.e. the Milky Way analogues) changes with variations to the stellar feedback parameters, by fitting to the values inferred from abundance matching. We find that there are significant degeneracies in the stellar feedback parameters when constraining the present-day stellar mass of the host, where the degeneracy corresponds to a freedom in the formation time of the stellar component. Additionally, at the end of Section~\ref{section:analysis}, we present the dependence of the number of luminous satellites on the variations in the stellar feedback, reionisation redshift and WDM mass. Finally, in Section~\ref{section:summary} we summarise our results and conclude.

\section{Simulation details} \label{section:sim_details}

Here we describe the key details of the simulations presented in this work. We begin by describing the aspects of the simulations and analysis that are constant throughout this work. This includes how the initial conditions are generated (Section~\ref{section:ICs}) and the details of the halo finder and merger tree (Section~\ref{section:halo_finder}). In Section~\ref{section:sim_details_varied} we focus on the parameters and associated routines that are varied and emulated in this work. This includes the stellar feedback, the star formation model, the reionisation redshift and the WDM particle mass.

\subsection{Initial conditions} \label{section:ICs}
All of the simulations share the same base $\Lambda$CDM cosmological parameters, using the WMAP9 best fit values \citep{WMAP9}. Specifically, $H_{0} = 70$kms$^{-1}$Mpc$^{-1}$, $\Omega_{\rm{m}}=0.2793$, $\Omega_{\rm{b}}=0.0463$, $\sigma_8 = 0.8211$ and $n_{\rm{s}} = 0.972$. The initial conditions are generated at $z=127$ using the \texttt{CAMB} \citep{camb} predicted $\Lambda$CDM linear power spectrum, which is then modified for the given WDM mass (see Section~\ref{section:wdm}).

To generate the zoom-in initial conditions we use \texttt{MUSIC} \citep{Hahn_2011}, with separate transfer functions for the DM and baryons. Systems in the original ARTEMIS sample were identified for resimulation by first running $25$ Mpc$/h$ box with $256^3$ collisionless particles. From this, haloes were identified in the mass range $8 \times 10^{11} < M_{\rm{200c}}/M_{\odot} < 2 \times 10^{12}$, to bracket the current uncertainty in the Milky Way's mass from a variety of observations \citep[e.g.][]{Guo_10,Deason_12,McMillan_17,Watkins_19, Callingham_19, Wang_2020}. The Lagrangian regions to resimulate were identified to contain all particles within $2R_{\rm{200c}}$ at $z=0$. The high resolution zoom-in region uses a DM particle mass of $1.17 \times 10^5$ $h^{-1}$$\rm{M_{\odot}}$ and an initial gas mass of $2.23 \times 10 ^4$ $h^{-1}$$\rm{M_{\odot}}$.

The original ARTEMIS sample was selected solely on halo mass, with no additional cuts based on isolation or formation history. Therefore, the sample (now constituting $45$ systems) is representative of haloes that form at this mass scale, with the caveat that the original simulation volume was $25$ Mpc$/h$. As such, particularly rare environments, such as large galaxy clusters, are not sampled.

From the original sample we focus on resimulating $3$ haloes. These were again selected based on present day halo mass (chosen to cover the sampled mass range), with no explicit selection on formation history or isolation. Using the naming convention from the original paper, these are haloes G42, G19 and G44\footnote{G44 was not part of the original sample of 42 haloes in \citet{Font_2020}, but was subsequently added to the sample in \citet{Font_2021}.} with halo masses of $M_{\rm{200c}} = 5.68 \times 10^{11}$, $9.18 \times 10^{11}$ and $1.32 \times 10^{12}$ $h^{-1}$$\rm{M_{\odot}}$ in the DM only simulation.

\subsection{Halo finder, merger trees and mass definitions} \label{section:halo_finder}

Collapsed, bound structures are identified using the \texttt{SUBFIND} halo finding algorithm, last described in \cite{Subfind}. Groups of haloes are initially identified using the friend-of-friends (FOF) algorithm, before individually bound structures within a given FOF group are identified using the \texttt{SUBFIND} algorithm. The most massive of these is then identified as the central, or host, while all other subhaloes are considered to be satellites. \texttt{SUBFIND} uniquely identifies individual particles as belonging to a given subhalo through an iterative unbinding algorithm.

Merger trees are generated using the \texttt{D-haloes} algorithm, using only the collisionless DM particles to track progenitors. The code is based on the algorithms of \cite{Srisawat_13} and \cite{Jiang_14}. In general, the algorithm uses the most bound particles of a given subhalo to track its progenitors and descendents. From this initial linking between snapshots the merger trees are then built, taking into account haloes missing in the \texttt{SUBFIND} catalogues at a given snapshot and may be linked to multiple later snapshots. See the previous references for details.

Throughout we will use various mass definitions. For total halo mass we use an overdensity definition such that the mean enclosed density is some multiple of the background density. For comparison with other works we primarily use the definition from \cite{Bryan_norman_98}, which for our assumed cosmology represents a density contrast of $\Delta \approx 98$ with respect to the critical density. For stellar mass we either use a fixed spherical aperture (primarily $30$kpc), or use all particles that are identified as being bound from the \texttt{SUBFIND} algorithm. Throughout the paper we will specify the particular mass definition used and, where appropriate, motivate its use.

\subsection{Parameters for baryonic physics and dark matter} \label{section:sim_details_varied}

All of the simulations use the \texttt{PGadget-3} code (last described in \citealt{Gadget}) with the hydrodynamics implementation and galaxy formation (subgrid) physics developed for the EAGLE project \citep[][]{Eagle,Eagle_cal}. The EAGLE model includes prescriptions for metal-dependent cooling in the presence of a photo-ionising UV background, star formation, stellar evolution and chemical evolution, black hole formation and growth, along with stellar and AGN feedback.

In this work we are interested in exploring the joint effect of baryonic (subgrid) processes and potential small scale cosmological extension on Milky Way mass systems and their satellite populations. Therefore, we restrict our analysis to variations of the baryonic processes that are most important for these mass scales. Specifically, we explore variations in the stellar feedback parameters, the density threshold for star formation, and the reionisation redshift. Here we describe how these processes are implemented in the EAGLE model, along with the associated subgrid parameters. All other subgrid routines and parameters use the fiducial values presented in the original EAGLE simulation \citep[see][for details]{Eagle,Eagle_cal}.\footnote{Specifically the EAGLE Recal-L025N0752 simulation.}

The simulations presented model the effects of AGN feedback, however the associated parameters are held fixed throughout. IN general, it is expected that AGN feedback is the dominant for high mass haloes, while stellar feedback dominated at lower masses with haloes of approximately Milky Way being the transition between these two regimes and being the most efficient at forming stars \citep[e.g.][]{Moster_13, Behroozi_13}. As such, it is expected that AGN play a sub-dominant role in the formation and evolution of Milky Way mass haloes for many observables, with the gas fractions being a notable exception \citep[e.g.][]{Croton_06,Bower_08,Booth&Schaye_09,Davies_2019}. While it would be interesting to explore potential changes to both stellar and AGN feedback, this would necessitate a much larger number of simulations to maintain the accuracy of the emulator. As such, we have chosen to focus on the most important parameters for systems of Milky Way mass and smaller (i.e. stellar feedback and reionisation), and hope to explore a joint variation of stellar and AGN feedback in the future.

\subsubsection{Warm dark matter} \label{section:wdm}
In this work we study warm dark matter (WDM) as an extension to the standard CDM model. In general, WDM models assume that DM consists of a light, as yet undiscovered, particle that is relativistic in the early universe. These non-negligible initial velocities allow for DM to free stream, leading to the suppression of density fluctuations and structures on small scales. Assuming a given particle physics model, the physical scale that these suppression occur on can be interpreted as a particle mass. In practical terms within the simulations WDM results as a change to the initial conditions, which can be described through the linear power spectrum.

The linear power spectrum for a WDM cosmology can be written as transfer function, $T_{\rm{WDM}}$, with respect to a ($\Lambda$)CDM power spectrum counterpart,
\begin{equation}\label{eqn:wdm_transfer}
	P_{\rm{WDM}}(k) = T^2_{\rm{WDM}}(k) P_{\rm{CDM}}(k).
\end{equation}
Here we use the fitting function of \cite{Bode_2001}:
\begin{equation}\label{eqn:wdm_suppression}
	T_{\rm{WDM}}(k) = [1+(\alpha k )^{2 \nu}]^{-5/\nu}.
\end{equation}
Here $\nu$ represents the form of the cut off and $\alpha$ the corresponding scale of the cut off. The values used correspond to the best fit parameters from \cite{Viel_2005}. Specifically $\nu=1.12$ and
\begin{equation} \label{eqn:wdm_suprresion_fit}
	\alpha = 0.049 \bigg ( \frac{m_{\rm{DM}}}{1 \textrm{keV}} \bigg ) ^{-1.11} \bigg ( \frac{\Omega_{\rm{DM}}}{0.25} \bigg ) ^{0.11} \bigg ( \frac{h}{0.7} \bigg ) ^{1.22} h^{-1}\textrm{Mpc} .
\end{equation}
It is then the assumed WDM particle mass, $m_{\rm{DM}}$, that is varied. $\Omega_{\rm{DM}}$ is the cosmic fraction of DM, which is held fixed in this work to the WMAP9 best fit value, $\Omega_{\rm{DM}} = 0.233$ \citep{WMAP9}. The values used above, and relation to DM particle mass, assume that WDM is made of thermal relics. However, as the key change to the growth of structure in WDM simulations is the suppression in the initial matter power spectrum, this can effectively mimic other WDM models such as sterile neutrinos \citep[e.g.][]{Dodelson_94,Shi_99} and, to a more limited extent, cosmological extensions with a similar suppression, such as fuzzy DM \citep[e.g.][]{axion,Mocz}. All other cosmological parameters, such as $\Omega_{\rm{m,0}}$ and $H_{\rm{0}}$, are fixed to the values presented in the previous section.

The technical details of generating the zoom-in initial conditions are the same as described in Section~\ref{section:ICs}, with the $\Lambda$CDM initial power spectrum generated using \texttt{CAMB} and modified according to the above equations.

\subsubsection{Star formation threshold}

Star formation in the EAGLE model follows the pressure law scheme introduced in \cite{Schaye_Vecchia_08}, where it was shown that the observed Kennicutt-Schmidt law \citep{Kennicutt_98} can be converted to a relation between the star formation rate and the pressure of the gas in the simulations, given an assumed equation of state and under the approximation that the gas is self gravitating. The advantage of this scheme is that the observed parameters for the Kennicutt-Schmidt law (i.e. the slope and normalisation) can be explicitly specified as input parameters to the simulations. In this work we use the same values presented in the original EAGLE project.

Star formation only occurs in cold, dense gas. In EAGLE, star formation is regulated by a density threshold, $n^*_{\rm{H}}$, above which gas follows the pressure law scheme described above. The EAGLE model uses a metallicity dependent threshold originally proposed by \cite{Schaye_04},
\begin{equation} \label{eqn:star_form_thresh}
	n^*_{\rm{H}} = \rm{min} \bigg [ n^*_{\rm{H,0}} \bigg (\frac{Z}{0.002} \bigg ) ^{-0.64} , 10 cm^{-3} \bigg ].
\end{equation}
The general form of the metallicity dependence is motivated in \cite{Schaye_04}, while the maximum value is specified to prevent arbitrary large density thresholds in low metallicity gas.

Both \cite{Schaye_04} and the original EAGLE simulations use $n^*_{\rm{H,0}}=0.1 \rm{cm}^{-3}$. Within the simulations the density threshold represents the transition at which the cold phase of gas (which simulations typically cannot resolve directly), is expected form. Typically,  this threshold cannot be observed directly, and instead is indirectly constrained from the observed star formation rates of nearby disk galaxies. Due to the theoretical uncertainties in deriving such a threshold, the diverse range used in current simulations as well as the choice of density threshold having a significant effect on dwarf galaxies \citep[e.g.][]{Benitez_19}, we choose to vary $n^*_{\rm{H,0}}$.

\subsubsection{Stellar feedback}

The EAGLE model uses the stochastic thermal feedback prescription originally presented in \cite{Vecchia_schaye_2012} to model the effects of Type II supernovae. Each star particle has a chance of undergoing a feedback event where neighbouring gas elements are instantaneously heated by a fixed temperature increment, $\Delta T _{\rm{SF}}$. The probability of such a feedback event occurring can be calculated from the given $\Delta T_{\rm{SF}}$ and available energy (see \citealt{Vecchia_schaye_2012} for details). Typically, the energy available for stellar feedback from a Type II supernova is taken to be $1.736\times 10^{49}$ $\rm{erg \, M}^{-1}_{\odot}$, assuming a Chabrier \citep{Chabrier2003} initial mass function. However, there is freedom within the model to allow a certain fraction, $f_{\rm{th}}$, of this fiducial energy to couple to the surrounding gas. The freedom in $f_{\rm{th}}$ was used to calibrate the original EAGLE and ARTEMIS simulations, and is therefore a key focus in this work.

\begin{figure}
	\centering
	\includegraphics[width=0.48\textwidth]{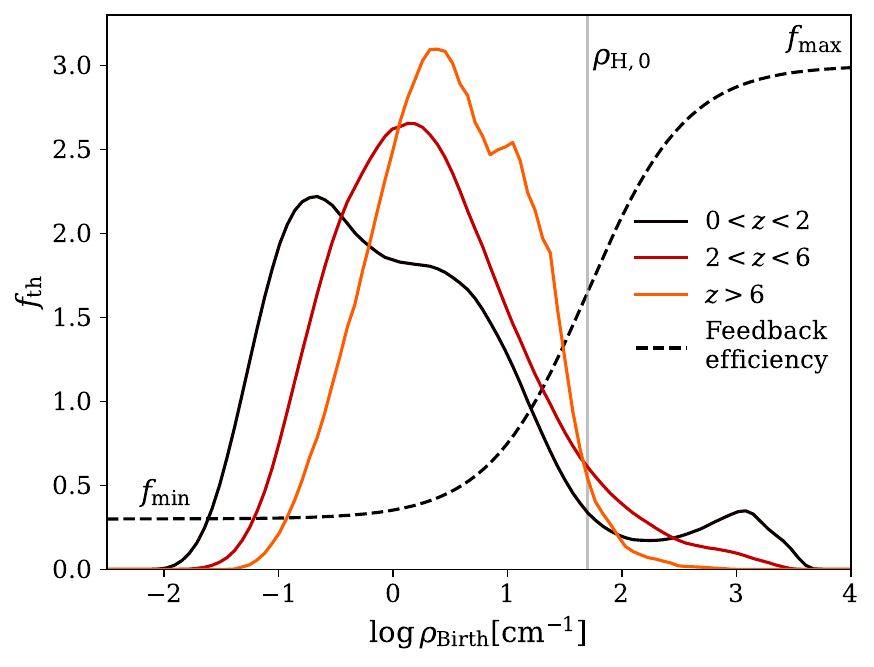}
	\caption{The black dashed line shows the dependence of the stellar feedback efficiency parameter, $f_{\rm{th}}$, on the stellar birth density. The plotted dependence corresponds to the choice of parameters used for the original ARTEMIS suite, $f_{\rm{max}} = 3$, $f_{\rm{min}} = 0.3$ and $\rho_{\rm{H,0}} = 50 \rm{cm}^{-1}$ (see Eqn.~\ref{eqn:stellar_efficency} for definitions). Additionally plotted for reference is the present day normalised distribution of stellar birth densities for all bound star particles of halo G42, split into bins according to their birth redshift (see legend).}
	\label{fig:stellar_feedback}
\end{figure}

In the EAGLE model $f_{\rm{th}}$ is allowed to vary as function of the star particle's birth density, $\rho_{\rm{H,birth}}$, with the following parametric relation,
\begin{equation}\label{eqn:stellar_efficency}
	f_{\rm{th}} (\rho_{\rm{H,Birth}})= f_{\rm{min}}+\frac{f_{\rm{max}}-f_{\rm{min}}}{1+\big(\frac{\rho_{\rm{H,birth}}}{\rho_{\rm{H,0}}}\big)^{-\alpha}}.
\end{equation}
The form of the above relation leads to more energy being coupled gas in denser environments, i.e. larger value of $f_{\rm{th}}$ at higher values of $\rho_{\rm{Birth}}$, and vice versa.\footnote{The EAGLE simulations also implemented a metallicity dependence to $f_{\rm th}$ that we do not include here. However, the dominant effect is due to the density dependence, as shown in figure~3 of \cite{Eagle_cal}.} The general behaviour of the relation is designed to compensate for feedback events being numerically inefficient at heating high density gas, for which the stellar birth density is used as a proxy. The relation between $f_{\rm{th}}$ and $\rho_{\rm{H,Birth}}$ is shown in Fig.~\ref{fig:stellar_feedback} as the dashed black line. In general the relation between $f_{\rm{th}}$ and $\rho_{\rm{H,birth}}$ resembles that of a smoothed step function. $f_{\rm{min}}$ corresponds to the minimum efficiency at small densities, $f_{\rm{max}}$ the maximum at high densities, while $\rho_{\rm{H,0}}$ controls the transition scale between the two regimes and $\alpha$ controls how quickly the transition occurs.

The values used in the original EAGLE simulation (specifically, the EAGLE Recal-L025N0752 simulations) were $~{f_{\rm{min}}=0.3}$, $~{f_{\rm{min}}=3}$, ~{$\rho_{\rm{H,0}} = 10$cm$^{-1}$} and $~{\alpha=1}$. In the ARTEMIS simulations, which have a particle mass resolution $8$ times higher than EAGLE Recal-L025N752, the stellar feedback was recalibrated, using $\rho_{\rm{H,0}} = 50$cm$^{-1}$, to better fit the present-day stellar mass halo mass (SMHM) relation at the Milky Way mass scale. Fig.~\ref{fig:stellar_feedback} shows the dependence of the stellar feedback efficiency parameter, $f_{\rm{th}}$, on the stellar birth density with values assumed in ARTEMIS (black dashed line).

To further explore the freedom in matching the observables within the stellar (subgrid) routine described above, we choose to simultaneously vary $f_{\rm{min}}$, $f_{\rm{max}}$ and $\rho_{\rm{H,0}}$. We find it more useful to express $f_{\rm{min}}$ as a fraction of $f_{\rm{max}}$, specifically
\begin{equation} \label{eqn:stellar_effiency_fmin}
	f_{\rm{min}} = A f_{\rm{max}},
\end{equation}
where $A$ is then the emulated parameter (rather than $f_{\rm{min}}$). This mild reformulation has a few distinct advantages. It is much easier to ensure that $f_{\rm{max}}>f_{\rm{min}}$ (corresponding to $A<1$), as well as being more intuitive to present the stellar feedback efficiencies in a relative manner rather than as absolute values. Throughout this work we fix the slope of the transition $\alpha$ to a value of 1 (i.e. we do not emulate this parameter). During the development of this project it was found that $\alpha$ has a minimal effect on the results\footnote{In our analysis we only considered $\alpha>0.5$. It is likely that very small choices of $\alpha$ would result in noticeable differences.}.

In summary, we emulate the effects of three parameters associated with stellar feedback in the EAGLE model, $f_{\rm{max}}$, $A$ and $\rho_{\rm{H,0}}$. This allows for the relation between the stellar efficiency, $f_{\rm{th}}$, and the star particle's birth density to be systematically varied.

Additionally plotted in Fig.~\ref{fig:stellar_feedback} is the distribution of stellar birth densities for halo G42 from the fiducial (original) ARTEMIS simulations, selecting all star particles identified as bound to the host at $z=0$. These are then split into three bins according to the formation redshift of the star particles (see legend). The overall form of the relation is such that no stars are born in very low density environments ($\log \rho_{\rm{birth}} \lesssim -2$) due to the star formation threshold, while most stars form at intermediate densities. It can also be observed that the minimum birth density increases at higher redshifts. This is due to the metallicity dependent star formation threshold used (see the previous subsection for details), which allows the gas to form in less dense environments as gas becomes more enriched over time. The highest densities are additionally suppressed, this being directly related to the form of $f_{\rm{th}}$. The increase in the stellar feedback efficiency at high birth densities leads to a suppression of star formation in these regimes. If a constant feedback efficiency were used instead, the sharp decrease in the number of stars forming in high densities would not exist (see, for example, figure~7 of \citealt{Eagle_cal}).

The metallicity dependence for the star formation threshold described above explains the redshift evolution in the low $\rho_{\rm{H,birth}}$ regime in Fig.~\ref{fig:stellar_feedback}. In general, the metallicity of gas within the simulation will increase over time. As such, the star formation threshold will be comparably larger at high redshifts compared to today. It is therefore expected that the observed minimum birth densities of the stars will decrease with time, as shown in Fig.~\ref{fig:stellar_feedback}.

\subsubsection{Reionisation}

Radiative processes are modelled as a function of gas density, temperature and redshift by interpolating precomputed cooling tables using the CLOUDY model \citep{Cloudy}. Importantly for this work, the effect of reionisation is also implemented, following the scheme presented in \cite{Wiersma_09}. This includes HI reionisation that occurs instantaneously at a specified redshift, $z_{\rm{reion}}$. The original EAGLE (and ARTEMIS) simulations used $z_{\rm{reion}}=11.5$, consistent with Planck measurements at the time \citep{Planck_13}. Estimates for the reionisation redshift have since been re-evaluated, with most constraints suggesting a lower value of $z_{\rm{reion}} \sim 6$--$7$ \citep[e.g.][]{Robertson_15,Bouwens_15,Planck_18}. While reionisation is modelled to be instantaneous in the simulations, in reality, it is likely to happen over an extended time. This is supported by observations using different probes that are sensitive to different phases of the Universe's reionisation history. This provides further motivation for us to explore variations in the redshift of reionisation, $z_{\rm{reion}}$. By emulating this parameter, we can further understand the role   reionisation plays on the formation of the smallest galaxies (in the stellar mass regime $M_{\rm{stel}} \lesssim 10^5 M_{\odot}$), which are typically the most affected by these changes.

\section{Emulation} \label{section:emulator}

As is common throughout the field, we will use the term `emulator' to refer to a numerical scheme that allows for a fast prediction of the results from a (hydrodynamical) N-body simulation as a function of specified input parameters. In general, it is not possible to output an exact replica of a cosmological simulation (i.e. a list of all particle types and their properties). We aim instead to predict a range of summary statistics, $\vec{S}$. Examples of these include the stellar mass of the main galaxy, the number of satellites of a given mass, or any robust statistic that can be measured directly from the simulations. The goal of the emulator is then to predict these summary statistics as a function of the key input parameters, $\vec{\theta}$. In this work we use $6$ key input parameters, specifically ${\vec{\theta} = (m_{\rm{DM}}, A, f_{\rm{max}}, \rho_{\rm{H,0}}, n^*_{\rm{H,0}}, z_{\rm{reion}})}$. See Section~\ref{section:sim_details_varied} for definitions and descriptions of these parameters.

One limitation of the above `emulation' approach is that the summary statistics must first be specified. As such, the most powerful way of constraining the simulations may be missed. While in this present work we focus on emulating a range of summary statistics, the simulations are well suited to develop more advanced machine learning methods such as deep learning, which has previously be proven to efficiently extract significant information from a wide range of astrophysical and cosmological data \citep[e.g.][]{Storrie-Lombardi_92,Lochner_16,Villaescusa_21,Nguyen_24}

There are two key steps to build the emulator. Firstly, the input parameter space, $\vec{\theta}$, must be sampled. From this initial sampling the summary statistics are then measured and a regression model is trained to make predictions at combinations of $\vec{\theta}$ that are not directly sampled with simulations. Here we sample $\vec{\theta}$ using a Latin hypercube and then build the regression model (i.e. interpolate) by using a Gaussian process. The accuracy of the emulator depends strongly on both the sampling and regression model used, which we discuss below.

\subsection{Emulator parameters and sampling} \label{section:emulator_paramters}

To sample the parameter space we use a 6-dimensional orthogonal Latin hypercube consisting of $25$ nodes (i.e. sampled points). A Latin hypercube results in a uniform, homogeneous and space filling sampling, minimising the distance between nodes and in turn maximising the accuracy of the emulator for a given number of sampled points. Standard convention is to define the Latin hypercube such that all points are sampled on the range $[0,1]$. From this, each dimension is then mapped to each of the emulated parameters. For the baryonic parameters sampled here ($A$, $f_{\rm{max}}$, $\rho_{\rm{H,0}}$, $n^*_{\rm{H,0}}$ and $z_{\rm{reion}}$), this mapping is either done linearly or logarithmically, such that only the desired range of parameters to be sampled needs to be specified. The DM particle masses, $m_{\rm{DM}}$, are also sampled (this is described below). The Latin hypercube coordinates are then multiplied and translated by the appropriate factors to sample the entire range. A summary of the chosen ranges is shown in Table~\ref{Table:emulation_parameters}, along with the type of sampling (i.e. linear or logarithmic).

We note that it is difficult to know \textit{a priori} the correct range to sample for the variety of these parameters. For parameters with a clear physical analogue that can be measured from other observations, the choice is relatively clear, as the current (conservative) observational constraints should be covered. However, for parameters that are specific to the simulations, and which do not have a clear physical analogue that can be measured, it is not so clear what a reasonable sampled range should be. Ideally, the parameters should cover the observational uncertainties for galaxy properties of interest, however this range can often only be reliably derived by first having the emulator.

In this work the sampled ranges for the reionisation redshift, $z_{\rm{reion}}$, and the WDM mass, $m_{\rm{DM}}$, were chosen to conservatively cover the current observational constraints. The star formation threshold, $n_{\rm{H}}^*$ was chosen to sample up to a factor of $3$ from the fiducial value used in the EAGLE and original ARTEMIS simulations.

The ranges for $A$, $f_{\rm{max}}$ and $\rho_{\rm{H,0}}$ were chosen using an earlier version of the emulator trained on a narrower range of parameters. Here, the final ranges of these parameters were chosen to be the estimated (and extrapolated) $3\sigma$ constraint on each of these parameters when fitting the host stellar mass to the fiducial case, using only the emulator uncertainty. As discussed later (see Section~\ref{section:host_stellar_mass}, Fig.~\ref{fig:host_stellar_mcm}) we do find constraints on $A$ and $\rho_{\rm{H,0}}$ individually when fitting to the stellar mass, which suggests that this original estimation and extrapolation was driven by the earlier emulation range and the corresponding prior.

While the baryonic parameters are sampled in a relatively simple way, it is useful to sample $m_{\rm{DM}}$ in a more complex manner. Specifically we use the relation,
\begin{equation} \label{eqn:wdm_sampling}
	m_{\rm{DM}} =
	\begin{cases}
	-\frac{40}{7} x +\frac{47}{7} & \text{, $x>0.3$} \\
	\frac{3}{2}\frac{1}{x} & \text{, $x \leq 0.3$}
	\end{cases}
\end{equation}
where $x$ is assumed to be some uniform sampling in the range $[0,1]$ (as given by a Latin hypercube). It is desirable that the emulator, and in turn the chosen sampling, is able to reproduce the mass of the cold dark matter (CDM) particle exactly, which in this work we take it to be $m= \infty$.\footnote{In ($\Lambda$)CDM models, potential DM candidates are expected to be have particle masses $\sim$ GeV-TeV, where the suppression of the linear power spectrum happens well below the resolution limit of our simulations. Thus, for practical purposes, it is sufficient to treat $m= \infty$ for the ($\Lambda$)CDM case.} However, as any emulation range and its sampling must be finite, it is not possible to sample CDM using either a linear or logarithmic sampling of $m_{\rm{DM}}$. The above relation (Eqn.~\ref{eqn:wdm_sampling}) aims to address this issue, while allowing control of the sampling and accuracy of cosmologies close to the CDM case. The piecewise function consists of a combination of a linear sampling at small particle masses with a $1/x$ sampling for larger masses. This contraction at larger masses allows for the mass of CDM particles to be exactly sampled, where $m_{\rm{DM}} = \infty$ corresponds to $x=0$. The exact coefficients were chosen with two key mass scales in mind; the minimum sampled particle mass is $m_{\rm{DM}}=1$~keV, while $m_{\rm{DM}}=5$~keV represents the transition from the two sampling, it was additionally chosen so that $30\%$ of the sampled nodes correspond to $m_{\rm{DM}}<5$~ keV. The general motivation for these specific coefficients was to identify a WDM particle mass scale at which the effects of WDM begin to have a limited impact on the resolved haloes in our simulations, chosen to be $m_{\rm{DM}}=5$~keV.

\begin{figure*}
	\centering
	\includegraphics[width=\textwidth]{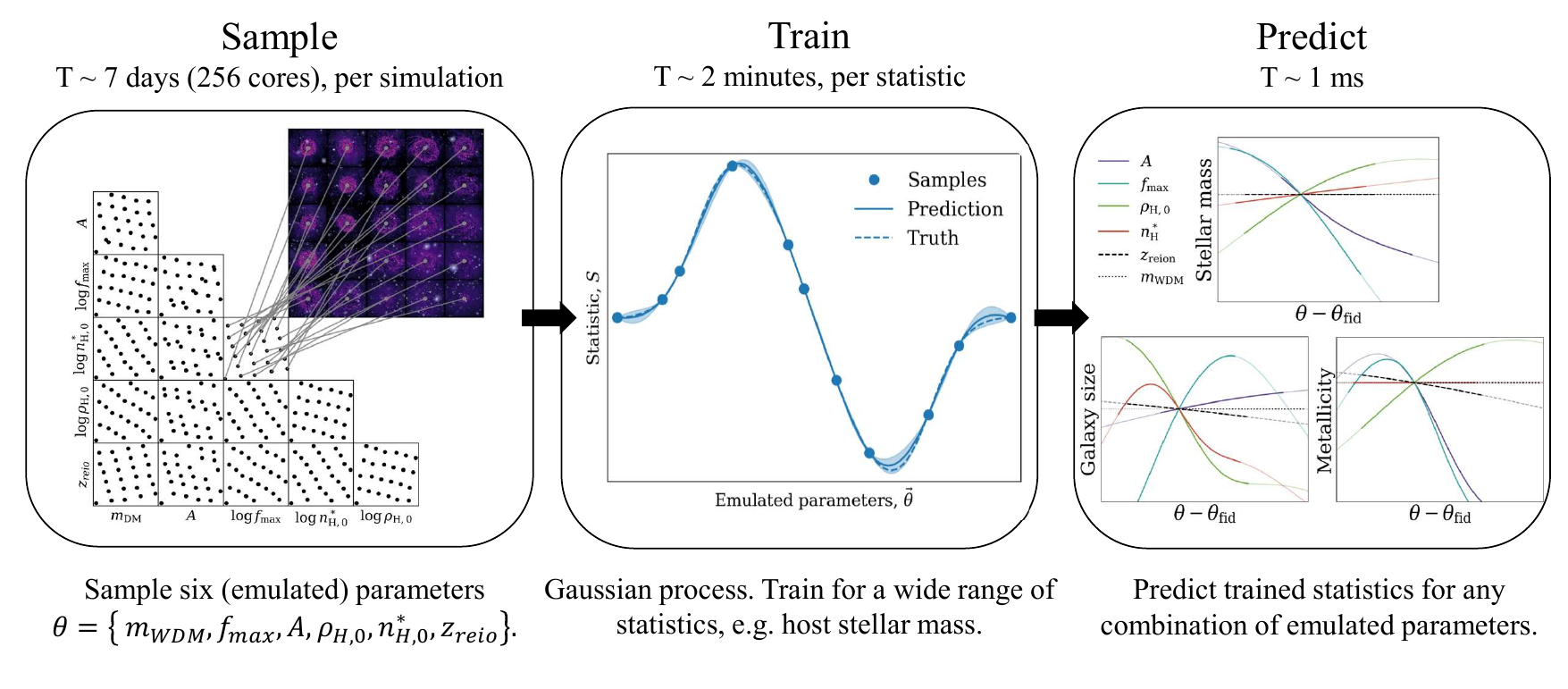}
	\caption{Schematic summary of how the emulators are built. First, the available parameter space is sampled with simulations (left panel). From these, many Gaussian processes are trained for a wide range of statistics  (middle panel), then finally the emulator is used to predict these statistics at any combination of parameters within the sampled space (right panel). Additionally plotted above each panel are the approximate computing times for each of these steps, with the emulator offering a $\sim 10^{11}$ increase in computational speed.}
	\label{fig:emulator_summary}
\end{figure*}

A summary for the six emulated parameters, along with the equations defining them, the fiducial values used in the original ARTEMIS simulations and their range of values sampled is given in Table~\ref{Table:emulation_parameters}. The left panel of Fig.~\ref{fig:emulator_summary} shows all 2-dimensional projections of the Latin hypercube used in this work, where the smooth sampling can be observed. For each combination of parameters a simulation is then run for each of the three haloes. We additionally run $4$ random combinations of parameters as hold out tests to evaluate the accuracy of the emulator. In total $3 \times (25+4)=87$ separate simulations are presented in the main suite, with an additional $10$ used to evaluate the stochasticity of the simulations and measured galaxy properties.

\begin{figure*}
	\centering
	\includegraphics[width=\textwidth]{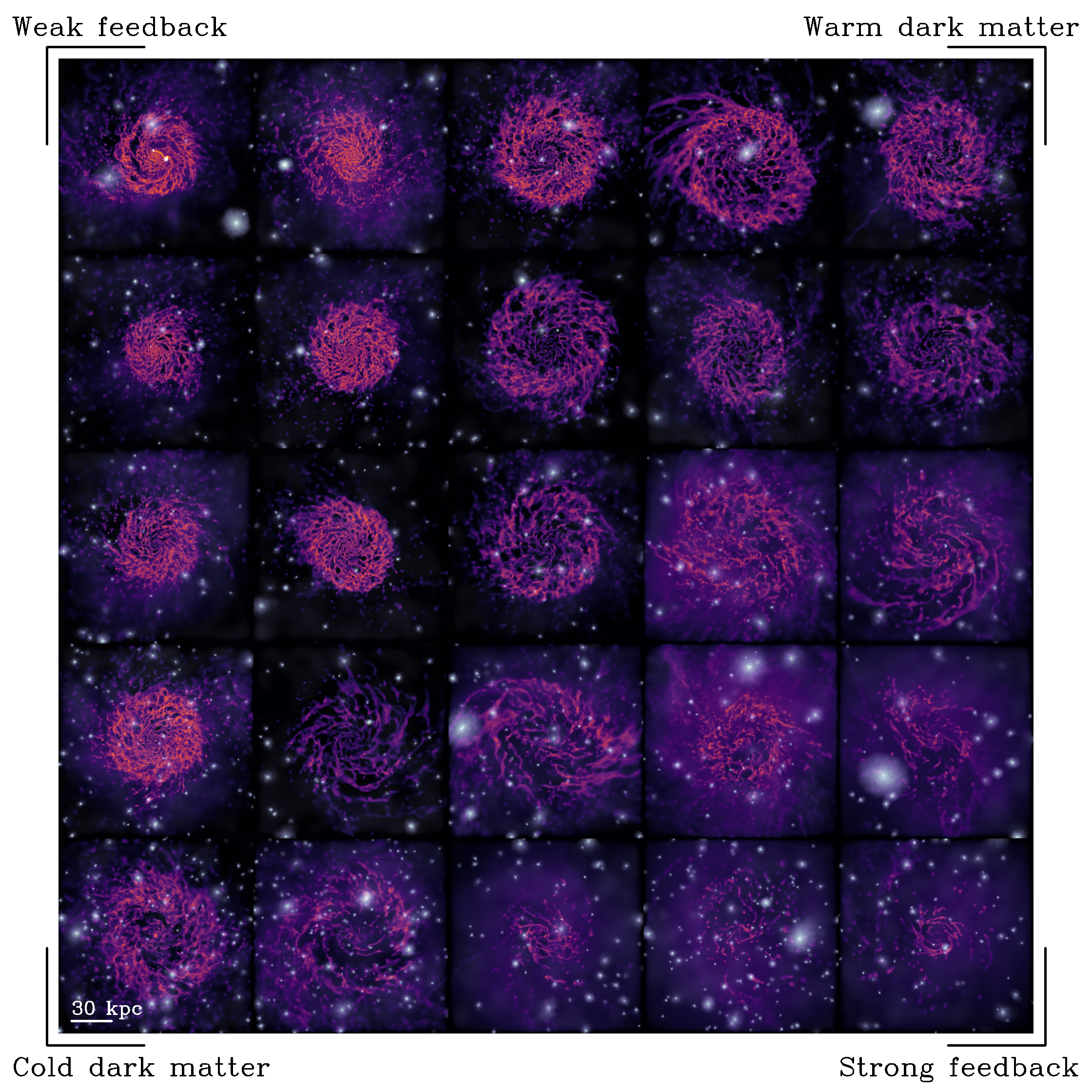}
	\caption{Visualisation of halo G42 for the $25$ sampled simulations, each with a different combination of stellar feedback parameters, star formation threshold, reionisation redshift and WDM mass. These simulations are used to build the emulators, and can be effectively treated as the training data. The visualisations represent a composite image of the gas (purple colours) and the DM (white colours) projected densities, calculated using Py-SPH viewer \citep{Py_sph_viewer}. For the DM density maps the central halo has been removed to highlight the satellite populations. The panels are organised so that the galaxy with the largest stellar masses are in the top left, and the smallest in the bottom right. Each bottom-left to top-right diagonal is additionally sorted in terms of the WDM mass such that the top-right panels are the strongest WDM models (i.e. smallest particle masses, $m_{\rm{DM}}$), while the models closest to CDM are in the bottom-left.}
	\label{fig:visualisation}
\end{figure*}

A visualisation of the resulting $25$ sampled simulations for halo G42 is shown in Fig.~\ref{fig:visualisation}. The image shows a composite of the gas and DM density. DM particles from the central halo have been removed to highlight the satellite population. The DM density is shown in white, while the gas uses the purple colour map. The plot is ordered so that the systems with the largest stellar mass are in the top left, and the smallest stellar masses are in the bottom right (the difference in stellar mass between the two most extreme simulations is $\sim 2$~dex). Each diagonal is additionally organised so that the bottom left corresponds to the coldest DM models, and the top right the warmest. While the stellar component is not shown in this image there are clear systematic changes in the distribution of the gas, both in density and morphology, that correlates with the stellar mass. For large stellar masses (top left), i.e. inefficient stellar feedback, there exists a relatively small, dense star forming a disk of gas. For smaller stellar masses (bottom right), corresponding to more efficient stellar feedback, much of the gas has been blown from the inner regions and is distributed within a gaseous halo, with little co-rotating gas in the form of a disk. There are also clear systematic differences in the number and mass distribution of satellites between different WDM particle masses, with a stronger WDM model leading to fewer satellites.  This visualisation demonstrates the diverse range of scenarios that is sampled by these simulations, and can in turn be sampled by the emulator.

\begin{table}
\caption{Summary of the $6$ emulated parameters varied in the simulations. The first column shows the given parameter, the second the equations where they are defined, the third and fourth columns show the fiducial values and emulation ranges for these parameters, and the final column shows the type of sampling used.}
\label{Table:emulation_parameters}
\begin{tabular}{lllll}
\hline
Parameter   & Equation  & Fiducial   & Emulator & Sampling   \\
   &   & value   & range & scheme   \\ \hline
$m_{\rm{DM}}$ {[}kev{]} 			  & Eqns.~(\ref{eqn:wdm_transfer}--\ref{eqn:wdm_suprresion_fit} )                                                                         			  & $\infty$   	  & $[1.0, \infty]$ & Eqn.~(\ref{eqn:wdm_sampling})             			  \\
$A$                     			  & Eqns.~(\ref{eqn:stellar_efficency}--\ref{eqn:stellar_effiency_fmin}) & $0.1$  		  & $[0,0.6]$ & Linear                   			  \\
$\log f_{\rm{max}}$    			  & Eqn.~(\ref{eqn:stellar_efficency})                                                     			  & $0.48$  & $[-0.30,1.14]$ & Log.	\\
$\log \rho_{\rm{H,0}}[\rm{cm}^{-3}]$   & Eqn.~(\ref{eqn:stellar_efficency})                                                     			  & $1.70$ & $[-0.075,4]$ & Log.               			  \\
$\log n^*_{\rm{H,0}}[\rm{cm}^{-3}]$ & Eqn.~(\ref{eqn:star_form_thresh})                                                     			  & $-1$  & $[-1.5,-0.52]$ & Log. \\
$z_{\rm{reion}}$        			  & --                                                                                                            			  & $11.5$ 		  & $[5,20]$ & Linear \\      			 
\hline
\end{tabular}
\end{table}

\subsection{Emulator prediction}

Another key aspect of the emulator is the regression model used. The aim is to effectively interpolate between the sampled points so that a given statistic can be predicted for any combination of emulated parameters, $\vec{\theta}$, within the sampled range. Here we choose to use a Gaussian process regression model. There are a number of key features provided by a Gaussian process that make it well suited to build emulators. In addition to providing a prediction for the value of the statistic at the choice of parameters, $S(\vec{\theta})$, a Gaussian process also provides the uncertainty in this prediction, which allows the uncertainty in the emulator to be incorporated in the statistical analysis. Gaussian processes also perform well in accuracy and scaling with sparsely sampled, high-dimensional data, therefore they are ideal for emulating cosmological simulation outputs. For example, in this work we sample a 6-dimensional space with only $25$ nodes (simulations), with a typical uncertainty and accuracy of $\approx 10\%$.

The Gaussian process used here consists of an anisotropic Mat\'ern kernel\footnote{For a Mat\'ern kernel, a smoothness parameter of $\nu=2.5$ corresponds to a twice differentiable function \citep[e.g.][]{Matern_kernel}.} and a white noise kernel. The associated hyperparameters are then optimised to maximise the likelihood for each statistic. The Mat\'ern kernel models the covariances between data points, allowing for predictions between nodes, while the white noise kernel accounts for any intrinsic noise in the data.

The middle panel of Fig.~\ref{fig:emulator_summary} shows an example of a Gaussian process regression model applied on a one-dimensional data set. Here the true function is shown with the dashed line, while the uneven samples (nodes) are shown as scatter points. A Gaussian process is then trained on these data, with the predictions of the model being shown with the solid line and with associated 1-sigma errors. The prediction of the Gaussian process resembles closely the true function, with places of where it deviates still being within the quoted errors. The behaviour of the uncertainties is generally intuitive; at locations that are directly sampled (the nodes) the uncertainty is zero, and the uncertainty remains small when close to these nodes, while the local maxima in the uncertainties occur in between nodes.

The example in Fig.~\ref{fig:emulator_summary} shows the basics of a Gaussian process regression model. The key differences for the emulators developed here, are that these are applied to a 6-dimensional parameter space (i.e. the emulated parameters are $\vec{\theta} = [m_{\rm{DM}}, A, f_{\rm{max}}, n_{\rm{H,0}}, n^*_{\rm{H,0}}, z_{\rm{reion}}]$) and rather than predicting a single statistic (observable), they can predict a wide range of these. Throughout our analysis, we are using independently trained Gaussian processes for each individual statistics. However, it is often useful and more intuitive to group these individual Gaussian processes into a single statistic. For example, to predict the stellar mass of the host as a function of redshift, each redshift is trained separately. However, it is useful to group all of these individual predictions into a `stellar mass' that can be predicted at any redshift. Similarly, predictions for secondary statistics are also made by training parameters separately. An example of these secondary statistics is the cumulative stellar mass function of satellite galaxies, where the number of satellites above each specified mass bin is trained and predicted separately. In this case, it is more natural to treat them collectively, as a single statistic. The total collection of all trained Gaussian processes is what we refer from now on as `the emulator'.

\subsubsection{Parameter inference and likelihood specification}

A key motivation to develop emulators is to use them to perform parameter inference. However, to do this robustly the likelihood must be specified, taking into account the uncertainties, and potential covariances, of the observed data. This, of course, will depend on the particular observations and data sets used. A relatively simple example of constructing the likelihood for the SMHM relation is given in Section~\ref{section:host_stellar_mass}.

Due to the way the emulators are constructed, in particular that we currently only predict statistics for three individual haloes, there are a number of key assumptions that will likely need to be made. Firstly, that the three haloes represent random, independent samples from an underlying distribution. While this distribution can in principle be as complex as needed, many statistics will be well approximated by a (multivariate) Gaussian, the mean and (co)variance of which can be specified from the observations being compared to (e.g. the particular galaxy catalogue). Alternatively, the original ARTEMIS sample, or similar simulations such as EAGLE, could be used to motivate the covariance of the data, and further test the ability of the simulations to reproduce the observations.

\subsection{Emulator summary}

Fig.~\ref{fig:emulator_summary} also includes a schematic summary of how the emulator is built. Initially, the parameter space is sampled using $25$ simulations for each of the $3$ haloes chosen from the ARTEMIS sample (this step is shown in the left panel). From these simulations, Gaussian processes are trained for a wide range of different statistics (see middle panel), including the properties of the hosts and of their satellites. This then allows for these statistics to be predicted for any combination of the emulated parameters, within the sampled range (see right panel). The top of each panel shows the approximate running times for each of these steps. As it can be seen in this figure, the emulator provides a significant improvement in the running time compared to simulations. While a typical simulation runs by $t\sim5$~$\rm{days} \sim 10^5~\rm{s}$ on a few hundred cores, the emulator takes $t \sim 1~\rm{ms}$ on a single core. The significant improvement in speed (by a factor of $\sim 10^{11}$) underscores the importance of building and using emulators for astrophysical problems. Specifically for studying the small scale structure tensions, the substantial reduction in the computational cost allows for a fast and thorough exploration of the multi-dimensional parameter space, in conjunction with the use of more sophisticated statistical analysis methods, such as Markov-Chain-Monte-Carlo (MCMC) sampling, which would not be possible by directly running simulations.

In Appendix~\ref{section:accuracy_appendix} we present an analysis of the intrinsic scatter within the simulations along with a test of the accuracy of our model compared with simulations and choices of parameters not used to develop the model. In general, we find that the emulators are $\approx 10$--$30 \%$ accurate, depending on the statistics that are being considered. It is observed that the intrinsic scatter within the simulations is typically $\sim 5 \%$ (for the stellar mass of the main halo) and mildly correlated with redshift.

\section{Initial analysis and Results} \label{section:analysis}

In this section we present initial results from the suite of simulations and corresponding emulators. We begin by studying the host stellar mass, a key property that is sensitive to the stellar feedback and the main statistic that was used to re-calibrate the original ARTEMIS simulations. We also explore what freedom there is in matching other host properties, such as the metallicities, sizes, in-situ fractions (i.e., the fraction of stars formed in the most massive progenitor of the host galaxy) and galaxy morphologies. Metallicities are studied both as averaged values for each host and as metallicity distribution functions (MDFs) of their stars. Finally, we study the effects that changes in the stellar feedback, reionisation redshift and WDM particle mass have on the stellar mass function of satellite galaxies.

\subsection{Host stellar mass} \label{section:host_stellar_mass}
In this subsection we explore how the stellar mass of a Milky Way-mass host system varies as a function of the emulated parameters. As previously mentioned, this is the main statistic used to re-calibrate the EAGLE model for the original ARTEMIS simulations. It is therefore useful to explore what freedom there is within this initial calibration step, and whether the choice of parameters was unique.

\begin{figure}
	\centering
	\includegraphics[width=0.45\textwidth]{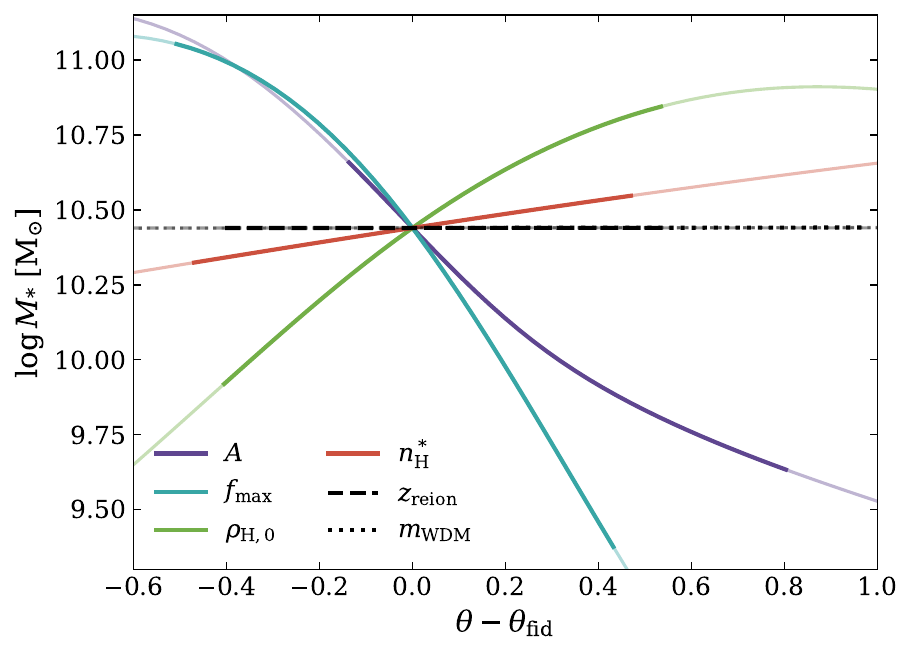}
	\caption{The dependence of the host stellar mass, defined as the mass within $30$~kpc, on the emulated parameters. Here each parameter is individually  varied (see legend), with the other five parameters held fixed to their fiducial values. The x-axis is in `emulator units', normalised such that the emulation range is from $0$ to $1$ and offset so that the fiducial choice is at the origin. Where the prediction is outside the emulators range the lines are plotted as transparent. The WDM mass and reionisation redshift have essentially no effect on the host stellar mass, while the star formation threshold has a mild effect over the sampled range, with the most important parameters being the three associated with stellar feedback, each able to affect the stellar mass by roughly an order of magnitude. The specific relations are shown for halo G42, with the other two systems showing very similar dependencies.}
	\label{fig:host_stellar_mass_param_dep}
\end{figure}

We start by studying how the stellar mass, computed within an aperture of $30$~kpc from the halo centre, changes when each parameter is varied individually. This is shown in Fig.~\ref{fig:host_stellar_mass_param_dep}, where each emulated parameter, $\theta$, is varied individually over its respective range, while the other parameters are held fixed to their fiducial values (see Table~\ref{Table:emulation_parameters}). This allows us to study the effect of each parameter variation in isolation. Later in this subsection, we will present an analysis where all parameters are allowed to vary simultaneously.

It is clear from Fig.~\ref{fig:host_stellar_mass_param_dep} that the host stellar mass is insensitive to both the assumed WDM mass, $m_{\rm{WDM}}$, and the reionisation redshift, $z_{\rm{reion}}$ (black dotted and dashed lines, respectively). This is consistent with other works for a halo with mass comparable to that of the Milky Way ($M_{\rm{200c}}\sim 10^{12}$ $M_{\odot}$), where it is expected that haloes of this mass should not be significantly affected by reionisation \citep[e.g.][]{Benson_2002,Wiersma_09} or by the suppression in density fluctuations for the range of WDM cosmologies with $m_{\rm{WDM}} > 1$~keV \citep[e.g.][]{Lovell_14,Bose_2016}.

The host stellar mass is mildly dependent on the star formation threshold, $n_{H,0}^*$, shown with a red line in this figure. Variations in the stellar mass are within $\approx 30\%$ of the fiducial value, across the entire range sampled in $n_{H,0}^*$. The relation here is positive, with larger density thresholds leading to an increased stellar mass for the host, which is consistent with results of other studies \citep[e.g.][]{Benitez_19}.

The most important parameters for setting the host stellar mass are found to be those associated with the stellar feedback efficiency, namely $A$, $f_{\rm{max}}$ and $\rho_{\rm{H,0}}$ (purple, blue and green lines). Each parameter can, in isolation, increase or decrease the stellar mass by roughly an order of magnitude from the fiducial case. The actual range of stellar masses able to be sampled is much larger ($10^{8.3} < M_{\ast}/\rm{M}_{\odot} < 10^{11.3}$) when the parameters are allowed to jointly vary. The relations are monotonic, with increases in $A$ and $f_{\rm{max}}$ leading to a decrease in the stellar mass, and increases in $\rho_{\rm{H,0}}$ resulting in an increase in stellar mass. The behaviour with respect to variations in $A$ and $f_{\rm{max}}$ can be understood by these parameters directly increasing (decreasing) the stellar feedback efficiency (see Fig.~\ref{fig:stellar_feedback} and Eqn.~\eqref{eqn:stellar_efficency}), resulting in less (more) star formation. The behaviour when $\rho_{\rm{H,0}}$ is varied can be readily understood from Fig.~\ref{fig:stellar_feedback}. Increasing $\rho_{\rm{H,0}}$ moves the transition from from low to high $f_{\rm{th}}$ values to a higher birth density, resulting in an overall decrease in the stellar feedback efficiency and in turn an increased stellar mass.

While individually varying the free parameters, as done above, is useful to build an intuition of the role of each parameter in isolation, we ideally want to explore the behaviour when all parameters are allowed to vary simultaneously, fitting to a given data set. We explore this for the host stellar mass, fitting to the stellar mass halo mass (SMHM) relation inferred from abundance matching. We restrict the following analysis to a CDM cosmology ($m_{\rm{DM}} = \infty$) and a fixed reionisation redshift of $z_{\rm reion}=11.5$, with both parameters having a negligible effect on the host stellar mass (see Fig.~\ref{fig:host_stellar_mass_param_dep}). We additionally only present the posteriors for the three stellar feedback parameters that are the most important for setting the stellar mass.

To fully explore the available parameter space, in this analysis 4-dimensions, we use an MCMC sampling. In a Bayesian framework the posterior on the parameters can be written, up to constant, as
\begin{equation}
p(\vec \theta | \vec x) \propto p(\theta) \times p(\vec x | \vec \theta),
\end{equation}
\noindent where $p(\vec \theta | \vec x)$ is the posterior on the free parameters, $p(\vec \theta)$ is the prior, and $p(\vec x | \vec \theta)$ is the likelihood. $\vec \theta$ represents the model parameters, and in this analysis there are only four free parameters: $f_{\rm{max}}$, $A$, $\rho_{\rm{H,0}}$ and $n_{\rm{H,0}}^{*}$. $\vec x$ represents the given data being fit to. Throughout, a flat prior with the same range as the emulator is used (see Section~\ref{section:emulator} for details).

To perform the MCMC analysis, we use the publicly available \texttt{python} package \texttt{emcee} \citep{emcee}. The MCMC sampling uses $32$ walkers with $50,000$ steps, initialised at the fiducial parameters used in the original ARTEMIS simulations (see Table~\ref{Table:emulation_parameters}), with an additional random $1\%$ scatter.

Here we fit the prediction from the emulator to the SMHM relation from \cite{Behroozi_19}. Assuming that the three haloes studied represent random, independent samples from the underlying SMHM relation, the likelihood can be written as
\begin{equation}
	\begin{aligned}
   \ln p(M_{\rm{\ast}} | \vec{\theta})=  \sum_{n} & -\frac{1}{2}\frac{[\log M_{\rm{\ast, obs}}(M_{\rm{vir, n}})-\log M_{\rm{\ast, pred,n}}(\vec{\theta})]^2}{\sigma_n^2(\vec{\theta})} \\ &+\ln [2\pi \sigma_n^2(\vec{\theta})],
   \end{aligned}
\end{equation}
\noindent where $\vec{\theta} = ( f_{\rm{max}}, \ A, \ \rho_{\rm{H,0}}, \ n_{\rm{H.0}}^{\ast} )$, and the sum is over all three haloes selected from the sample. $M_{\rm{obs,n}}$ is the observed average stellar mass for the given halo mass (taken from \citealt{Behroozi_19}), while $M_{\rm{pred,n}}$ is the stellar mass predicted for the given halo from the emulator. The halo mass, $M_{\rm{\rm{vir}}}$, uses the overdensity definition from \cite{Bryan_norman_98} and is measured from the DM-only simulations, for consistency with how the SMHM relation is derived in \cite{Behroozi_19}. This has the additional benefit of making the total halo mass, $M_{\rm{\rm{vir}}}$, independent of the choice of feedback parameters in this analysis. The error term, $\sigma_n$, is a combination of the intrinsic scatter in the SMHM relation, $\sigma_{\rm{scat}}$, and the uncertainty from the emulator, $\sigma_{\rm{em}} (\theta)$. These are assumed to be uncorrelated and added in quadrature,
\begin{equation}
	\sigma_n^2 (\vec{\theta}) = \sigma_{\rm{scat}}^2 + \sigma_{\rm{em,}n}^2 (\vec{\theta}).
\end{equation}
We assume $\sigma_{\rm{scat}} = 0.25$ dex, which is a value obtained by \citet{Behroozi_19} for the halo mass range sampled in our simulations. For reference, $\sigma_{n} \sim 0.1$ dex, although the value depends on the position in the emulator parameter space.

\begin{figure}
	\centering
	\includegraphics[width=0.485\textwidth]{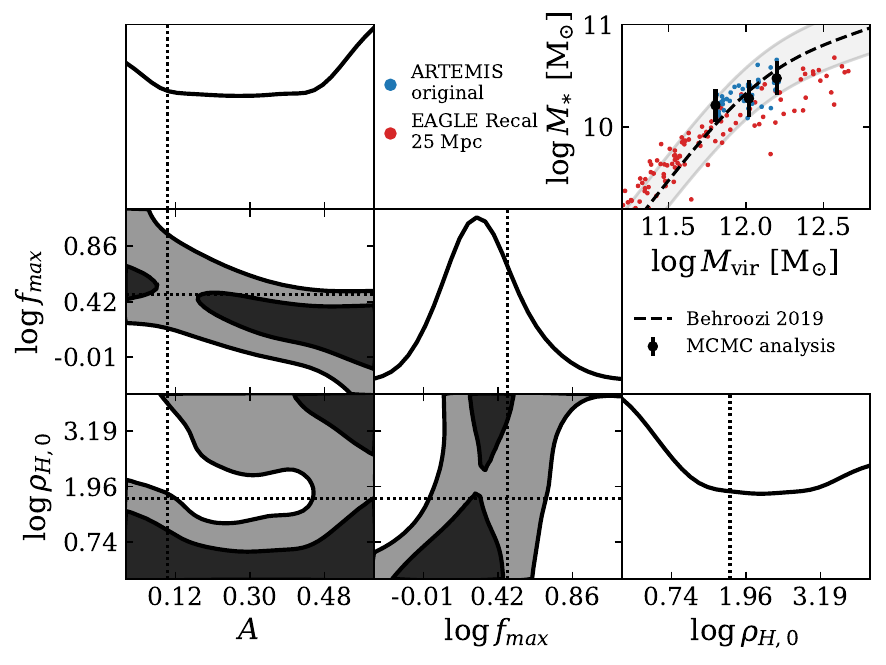}
	\caption{Top right panel shows the stellar mass halo mass relation, with $M_{\rm{*}}$ being the stellar mass with $30$kpc, while $M_{\rm{vir}}$ is the total halo mass. The \protect\cite{Behroozi_19} relation plotted as a dashed black line, the original ARTEMIS suite in blue, and the posterior of the MCMC analysis as black points. The points here show the mean and one sigma error bars. The bottom left panels show the corner plot of the MCMC posterior for the $3$ stellar feedback parameters, with the 1 and 2 sigma contours plotted. The black dashed lines show the fiducial combination of parameters. $\rho_{\rm{H,0}}$ is quoted in units of $\rm{cm}^{-3}$.}
	\label{fig:host_stellar_mcm}
\end{figure}

\begin{figure*}
	\centering
	\includegraphics[width=0.475\textwidth]{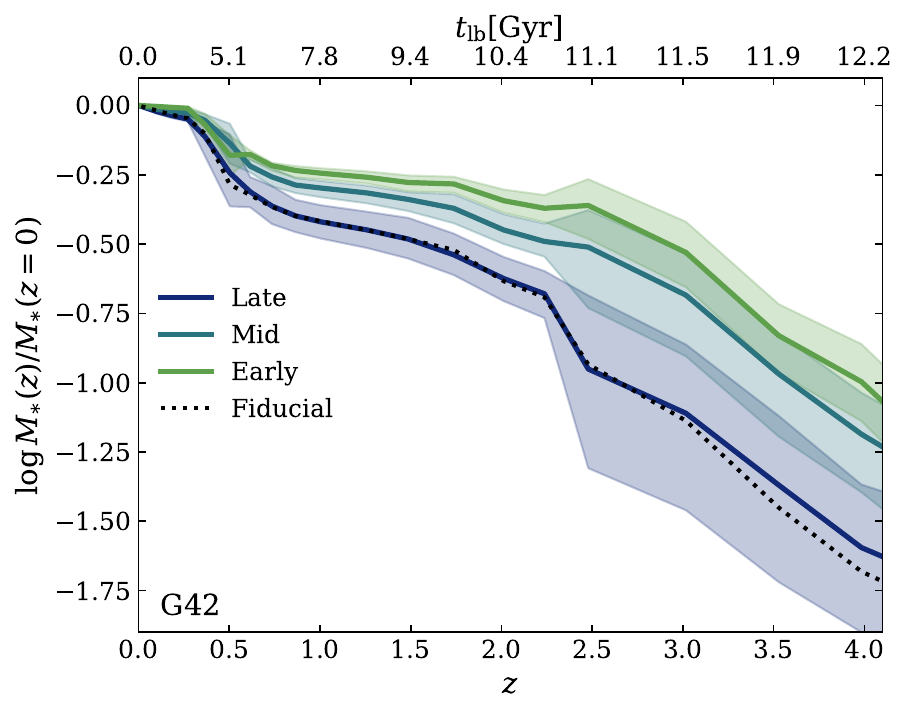}
	\includegraphics[width=0.465\textwidth]{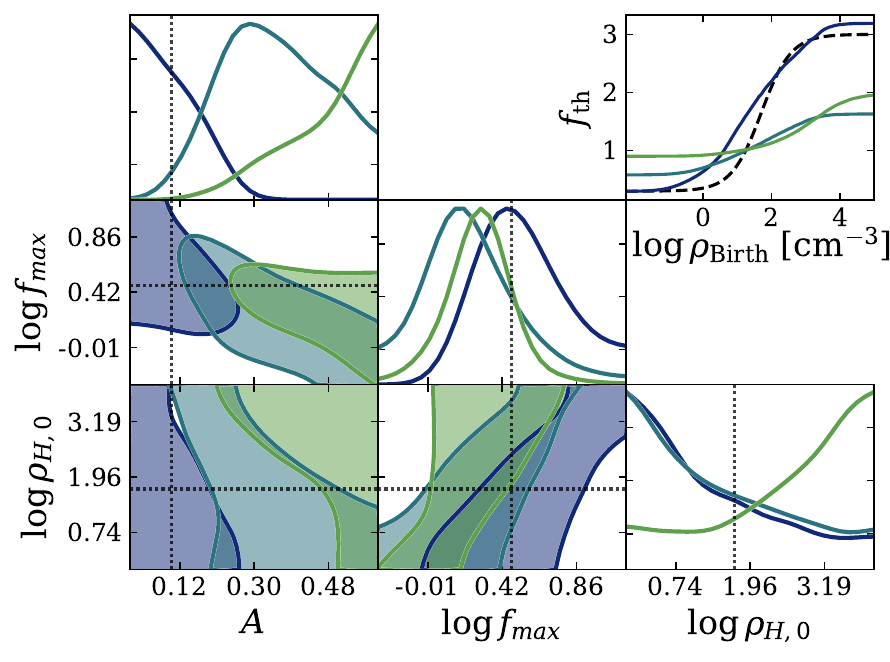}
	\caption{\textit{Left:} Evolution of the host stellar mass as a function of redshift for halo G42, normalised by the stellar mass today. All MCMC chains from Fig.~\ref{fig:host_stellar_mcm} are split into late, mid and early formation (see legend) according to being in the bottom, middle or top mean terciles at $z=2$ (see Section~\ref{section:host_stellar_mass} for details of the selection). Additionally plotted for comparison is the fiducial combination of parameters (dashed black line). Here this combination of parameters would be classed as late forming. \textit{Right:} Corner plot for the stellar feedback parameters (equivalent to Fig.~\ref{fig:host_stellar_mcm}) when split into the different formation scenarios. Here the 1-sigma contours are shown and the one-dimensional projections are normalised to their given maxima. There are clear systematic trends in the choice of parameters as a function of formation time, with $A$ showing the strongest correlation. The mean relation between the stellar feedback efficiency, $f_{\rm{th}}$, and stellar birth density, $\rho_{\rm{H, Birth}}$, for the three selections is shown in the top right panel. $\rho_{\rm{H,0}}$ is quoted in units of $\rm{cm}^{-3}$.}
	\label{fig:host_stellar_mcm_form_time}
\end{figure*}

The results of this MCMC analysis are shown in the Fig.~\ref{fig:host_stellar_mcm}. The top right panel shows the SMHM relation that is fit to, with the posterior of the MCMC chains shown as black error bars. As can be seen, it is a good fit to the data, matching closely the SMHM relation from \cite{Behroozi_19}. For reference, the original 45 Milky Way-mass haloes from ARTEMIS are plotted in blue, and the haloes from the EAGLE Recal simulation, shown in red. Both of these simulations match the SMHM by construction, with the ARTEMIS simulations having an additional recalibration for this statistic (see \citealt{Font_2020}). The posteriors for the three stellar feedback parameters are shown as corner plots in the bottom left panels, with added 1-$\sigma$ and 2-$\sigma$ contours. The dotted black lines in these panels are the fiducial combinations of parameters used in the original ARTEMIS simulations. Focusing initially on the one-dimensional posteriors, we see that there is little constraint on most of the parameters, with only $f_{\rm{max}}$ having a clearly preferred value. Both $A$ and $\rho_{\rm{H,0}}$ show a slight preference for choices at the edges of the emulation range. This is primarily due to the errors on the emulator being larger at the edge of the emulation range, rather than these parts of the parameter space offering a better fit to the data. We have explicitly verified this by evaluating the uncertainty of the emulator, $\sigma(\vec{\theta})$, at the edge of the sampled range. For parameters that are near the edge ($min(\vec{x})<0.05$ \& $max(\vec{x})<0.95$) the mean error is $\sigma = 0.14$ dex, while not near the edge ($0.05<max(\vec{x})<0.95$) the mean error is $\sigma = 0.11$ dex.

From the two-dimensional projections, it is clear that there are strong degeneracies between the three stellar feedback parameters. The existence of this degeneracy can be understood from the behaviour of the individual parameters (i.e. Fig.~\ref{fig:host_stellar_mass_param_dep}); for example, if a relatively large value of $A$ is used, which in isolation lowers the host stellar mass, then this can be compensated by decreasing $f_{\rm{max}}$ or by increasing $\rho_{H,0}$, both of these leading to an increase in the stellar mass. The three stellar feedback parameters can then work to compensate for each other. While strong degeneracies are present, there are still significant constraints on the parameter space. This is particularly clear where the parameters work in tandem to suppress or enhance star formation, such as when both $f_{\rm{max}}$ and $A$ have relatively large values. While this  behaviour is intuitive, it is so far only qualitative. To predict the exact, quantitative, form of the degeneracy we need to resort to the MCMC analysis, which in turn becomes possible from the results of the emulator.

 We also find that the degeneracy between the three stellar feedback parameters closely follows a surface in the three dimensions, as opposed to a single line. The $f_{\rm{max}}$--$A$ and $f_{\rm{max}}$--$\rho_{\rm{H,0}}$ projections view this surface relatively edge-on, while the $A$--$\rho_{\rm{H,0}}$ projection observes it close to face-on, resulting in the projected contours shown in Fig.~\ref{fig:host_stellar_mcm}. Using principal component analysis, the degeneracy surface can be well approximated by
\begin{equation}
	0.97 A + 0.25 \log f_{\rm{max}} -0.02 \log \rho_{\rm{H,0}} - 0.29 = 0,
\end{equation}
\noindent over the combined sampled ranges, subject to the condition $0 \leq A \leq 1$.

Having just seen that there are multiple combinations of the three stellar feedback parameters that lead to the same present-day stellar mass of the host, a natural next question is whether all of these feedback scenarios form galaxies with their final stellar mass in the same way. To answer this, we explore the redshift evolution of the host stellar mass, sampling the feedback parameters from the MCMC chains. We present this in the left hand panel of Fig.~\ref{fig:host_stellar_mcm_form_time}, where we present the stellar mass\footnote{Instead of using a fixed aperture to define the stellar mass, as done for the $z=0$ analysis, here we use all particles identified as bound by \texttt{SUBFIND}.}  as a function of redshift, normalised by the $z=0$ stellar mass. Here we show the fiducial combination of parameters (shown with black dashed lines) and the MCMC chains split into late, mid and early formation scenarios (which we describe shortly).

This figure indicates that there is significant freedom in the choice of feedback parameters when constrained to the present-day stellar mass. To further explore this, we choose to split the MCMC chains which all share the same stellar mass at $z=0$ (within the given uncertainties) into different formation scenarios. This is achieved by splitting the MCMC sample into terciles based on their stellar mass at $z=2$, which we refer to as late (bottom third), mid (middle third), and early (top third) scenarios. While this approach is straightforward on a halo by halo basis, ideally, we want the definition of an early, mid or late formation scenario to be unique for each MCMC chain. It is therefore necessary to average over all haloes. To do this, we calculate the percentiles for each MCMC chain prediction of the stellar mass at $z=2$ for each halo, and then average the values over all three haloes. This 'mean percentile', $P$, is then used to define a given MCMC chain as being a late, mid or early formation scenario, by applying the criteria $P>66$, $33<P<66$ and $P<33$, respectively.

In the left panel of Fig.~\ref{fig:host_stellar_mcm_form_time} the median stellar mass, with 1-$\sigma$ scatter, is plotted for these three formation scenarios. There is a clear separation between the three distributions. At $z=2$, this separation is by construction. However, the segregation appears at all redshifts, demonstrating that this selection does indeed define different formation times, and is not simply identifying noise within the data or a behaviour which is system specific. For reference, the stellar mass for the fiducial choice of parameters is also plotted as the dashed black line. Under this definition of formation time, the fiducial choice would be classed as `late' forming.

The distribution of feedback parameters ($f_{\rm{max}}$, $A$ and $\rho_{\rm{H,0}}$) split into the different formation times is shown in the right panel of Fig.~\ref{fig:host_stellar_mcm_form_time}, presented as a corner plot showing the 1-$\sigma$ contours. As it can be seen, the differences in formation times correspond to a systematic difference in the feedback parameters. This suggests that the freedom in the choice of parameters when constraining the present-day stellar mass directly corresponds to a freedom in choosing the formation time of the stellar component. Therefore, it is possible to choose both the present-day stellar mass, and the formation time with the appropriate combination of parameters. While all parameters separate more in their one-dimensional posteriors, compared to the total distribution, this most clearly happens for $A$. Generally, larger values of $A$ correspond to an earlier forming stellar component, and vice-versa. This behaviour, as well as $A$ exhibiting the most direct dependence on formation time, can be explained from Fig.~\ref{fig:stellar_feedback}. The dominant redshift evolution of the birth densities of stars happens at lower densities, with stars preferentially forming in lower density environments at later redshifts, while the number of stars that form at high densities is only mildly redshift dependent. Therefore, a higher value of $A$ corresponds to more efficient feedback at late times, which in turn would correspond to an early formation to result in the same stellar mass by $z=0$, as is enforced here. The redshift evolution appears to be predominantly controlled by $A$, with the other two parameters needed to be adjusted along the overall degeneracy to ensure the same stellar mass by $z=0$.

In the top right panel of the right hand corner plot we show the averaged relation between the feedback efficiency as a function of birth density for the MCMC chains split by formation time. This more clearly demonstrates the freedom that is allowed in this relation, and follows from the posterior of the feedback parameters. Here the fiducial combination of parameters (black dashed line) corresponds to the late formation scenario and represents a relatively large step (i.e. comparably large $A$). The two early formation scenarios then correspond to an overall smaller step between low and high $f_{\rm{th}}$, that is additionally shifted to higher birth densities. The three different formation scenarios separate most clearly at low $\rho_{\rm{H,birth}}$, which directly corresponds to $A$ being most clearly separated in the posterior.

\subsection{Complementary statistics}

\begin{figure}
	\centering
	\includegraphics[width=0.475\textwidth]{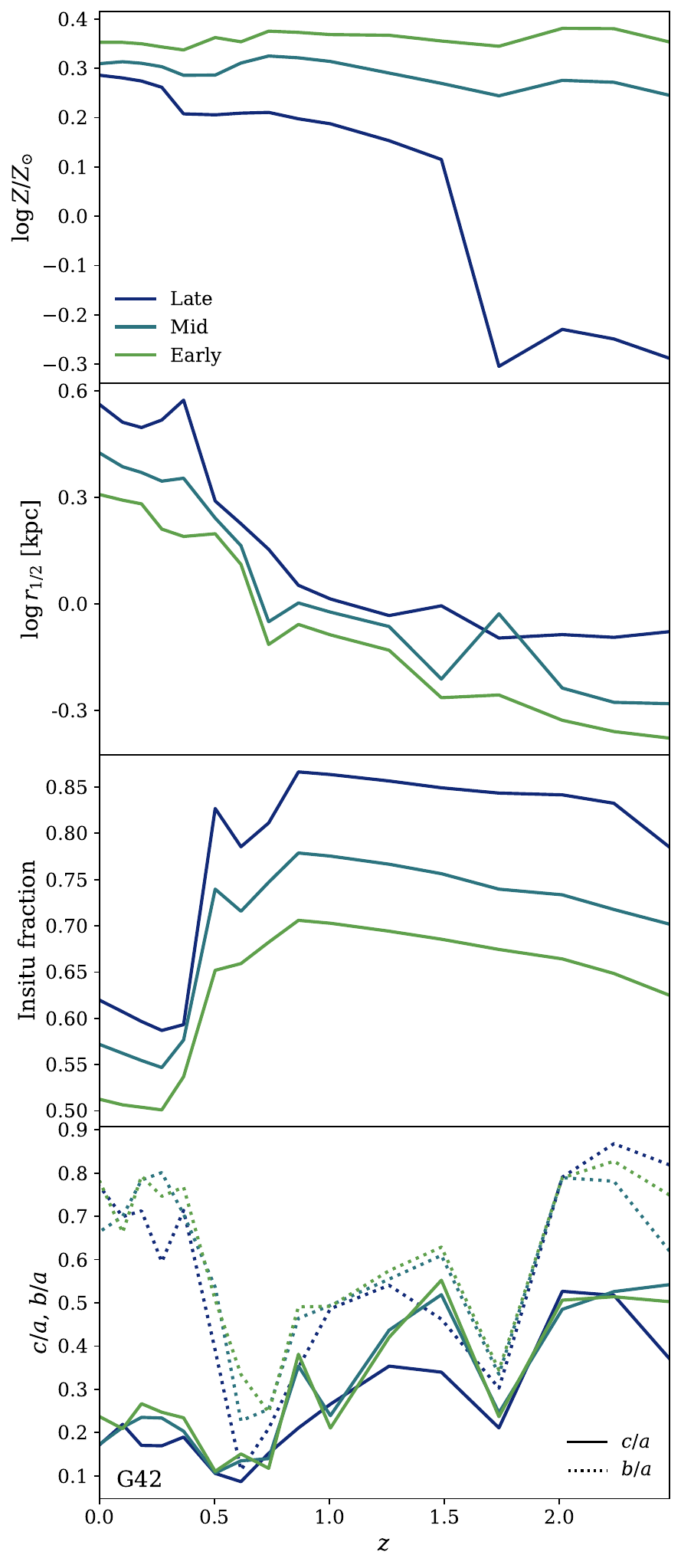}
    
	\caption{Redshift evolution of a range of statistics for the main progenitor for halo G42. The top panel shows the mass weight mean metallicity, the second panel the half mass radius, the third the in situ fractions and the fourth the stellar morphology, described by the ratio of the intermediate and/or minor to major eigenvalues of the moment of inertia tensor. The lines are averaged for the MCMC chains, which are constrained to have similar present-day stellar masses. These are then split into early, mid, and late stellar formation scenarios (see Fig.~\ref{fig:host_stellar_mcm_form_time}).}
	\label{fig:host_additional_stat}
\end{figure}

In the previous section, it was observed that there is a strong degeneracy in the stellar feedback parameters in setting the present-day stellar mass of the host. The freedom in the choices of feedback parameters corresponds to a freedom in the formation time of the stellar component. It is therefore interesting to consider if there are any other present-day galaxy properties that show systematic differences with stellar formation time, and can potentially be used to distinguish these choices of parameters. Here we focus on common statistics for the host galaxy, such as its size, metallicity and morphology, and properties sensitive to its formation history, such as the fraction of in situ and accreted stars.

In Fig.~\ref{fig:host_additional_stat} we present the redshift evolution of the main progenitor's metallicity, in situ stellar fractions, stellar half mass radius and morphology. All statistics are calculated from star particles identified as bound to the main progenitor. The metallicity is presented as the mass weighted mean metallicity, later we study the full metallicity distribution within the host.

The stellar morphology is described through the eigenvalues of the reduced moment of inertia tensor (calculated using the bound stellar particles). The specific form of the reduced moment of inertia tensor is
\begin{equation}
M_{i,j} = \sum_n \frac{m_{n} x_{i,n} x_{j,n}}{|\vec{x}_n|^2},
\end{equation}
where the sum is over all bound stellar particles, $m_{n}$ is the particle's mass and $\vec{x}_{n}$ its position. The major, intermediate and minor axes are then calculated from the square root of the eigenvalues. Here we present the ratio between the minor and major axes ($c/a$) and the intermediate and major axes ($b/a$). In this definition a disk corresponds to $c/a \approx 0$ and $b/a \approx 1$.

The final statistic we present here is the in situ vs ex situ fractions for the host galaxy. Here individual star particles are tagged as either being formed in situ or accreted. The procedure to make this identification is as follows. For each star particle we identify the time at which it was formed. We then track this particle in the snapshot after its formation. If the star particle at this redshift is identified as being bound to the main progenitor then it is tagged as forming in situ, otherwise it is identified as ex situ. This method follows the same procedure used in other papers using the ARTEMIS simulations \citep[e.g.][]{Font_2020}. There are many alternative methods used elsewhere in the literature, such as a stars birth radius from the main progenitor \citep[e.g.][]{Sanderson_2018} or methods to capture endo-debri \citep[e.g.][]{Cooper_2015}. However, in this work we are primarily interested in relative effects when using a consistent definition.

Focusing initially on the metallicity (top panel of Fig.~\ref{fig:host_additional_stat}), we see that, as with the stellar mass, the early, mid and late forming selections result in distinctly different redshift evolutions. The overall trend is as expected, with early star formation corresponding to a higher metallicity than late formation at higher redshifts, and vice-versa. Interestingly, while the high redshift ($z \gtrsim 2$) metallicities are distinct, these differences do not persist until the present day, with the different selections resulting in similar metallicities today. As such, it does not appear that the present-day metallicity is a powerful statistic in breaking the observed degeneracy in the stellar feedback parameters at this mass scale (see Fig.~\ref{fig:host_stellar_mcm}). If the present-day stellar mass is \textit{not} controlled for then there can be strong differences in the predicted metallicities, as shown shortly in Section~\ref{section:met_dist} (Fig.~\ref{fig:host_met_distribution}).  It therefore appears that the dominant factor in setting the present-day metallicity is the total amount of star formation, rather than when the stars are formed.

The stellar half-mass radius (second from top of Fig.~\ref{fig:host_additional_stat}) in general increases with redshift, as is expected for the galaxy, and halo, which are increasing in mass over these redshifts. Interestingly, there are clear trends (offsets) with formation time, which is relatively constant across all redshifts and is also seen for the other two haloes. Here we see that a scenario where the stellar component forms late results in a less concentrated distribution of stars than an early forming scenario. The difference in the stellar size is relatively constant with redshift, $\sim 0.3$ dex ($\sim 2$ kpc at $z=0$), notably persisting through to today.

Focusing next on the in situ fractions (third panel of Fig.~\ref{fig:host_additional_stat}), all scenarios have the same general form; at high redshifts the in situ fraction slowly increases with redshift, with the intrinsic star formation dominating over accretion, while at $z \sim 0.5$ there is a sharp decrease in the in situ fraction, before continuing to increase from $z=0.5 $ -- $0$. The particular form of the in situ evolution is unique to galaxy G42, that has a comparably high in situ fraction at early times and undergoes a significant merger at $z \sim 0.5$ resulting in a sharp decrease in the insitu fraction. The other two haloes. This can be seen in the evolution of $M_{\rm{\ast}}$ (Fig.~\ref{fig:host_stellar_mcm_form_time}). The other two haloes (G19 and G44) do not show such a clear feature in the evolution of the insitu fraction and have early values between $\approx 50 \%$ and $\approx 70 \%$ Here we also see a strong correlation with the formation time of the galaxy, with an early formation scenario resulting in a decreased in situ fraction, with a difference of $\approx 10\%$ over all redshifts. Significant differences in the in situ star formation and accreted populations offers a natural explanation of how there can be a significant change to the stellar evolution while the accretion history, in terms of DM haloes, is unchanged. However, the physical origin of this difference is not clear and is likely linked to the evolution of the stellar mass--halo mass relation in the dwarf regime. A full exploration of this is beyond the scope of this paper and will be the focus of future work.

Finally, we also explore the evolution of the morphology of the stellar component (bottom panel of Fig.~\ref{fig:host_additional_stat}), expressed through the eigenvalues of the moment of inertia tensor. Here the minor to major ratio, $c/a$, (solid lines) and the intermediate to major ratio, $b/a$, (dotted lines) are plotted. Using this definition a thin disk corresponds to $c/a \approx 0$, $b/a \approx 1$. Unlike the other statistics discussed here there is little to no clear correlation with formation time over all redshifts, with all lines broadly following each other. This particular galaxy has no obvious disk component until $z\sim 0.5$, where the merger appears to induce the formation of a stable disk. While the morphology is quite similar between the different formation times, when described through $b/a$ and $c/a$, the physical size of the galaxy has changed, meaning the disk height and size have in turn changed.

While Fig.~\ref{fig:host_additional_stat} shows the various statistics for galaxy G42, the other two systems show similar general trends. At high redshift, the metallicities are distinguishable between the different formation scenarios, however the $z=0$ metallicities are indistinguishable, with the other two galaxies in fact showing the late formation scenario having a slightly higher metallicity than the early scenario. The trends observed for $r_{1/2}$, the in situ/existu fractions, $Z$, $c/a$ and $b/a$ are qualitatively the same for all galaxies.

\subsubsection{Metallicity distributions} \label{section:met_dist}

In the previous section it was shown that the $z=0$ averaged metallicity of the central galaxy was broadly insensitive to the formation time of the stellar component. While the present-day averaged metallicities do not correlate strongly with the formation time (at fixed present-day stellar mass), it is possible that information is contained in the full metallicity distributions.

\begin{figure}
	\centering
	\includegraphics[width=0.485\textwidth]{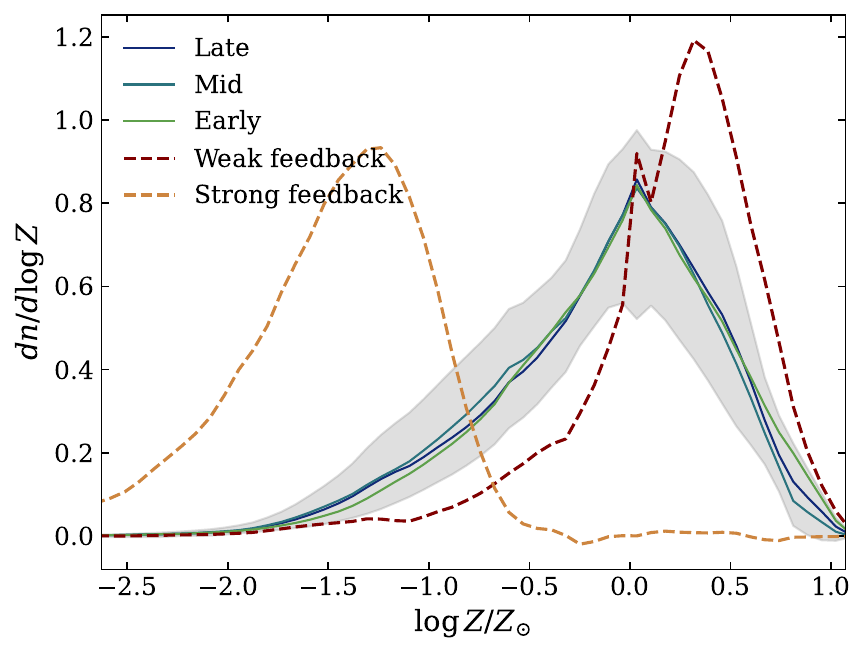}
	\caption{The mass weighted distribution of stellar metallicities for halo G42, with the integral normalised to unity. The solid coloured lines show the median average from the early, mid, and late formation scenarios (see legend). The grey band shows the one sigma scatter for the late scenario selection, with the two formation scenarios showing comparable scatter. For reference, weak and strong feedback cases are also plotted as dashed lines (see text for specific feedback parameters), that predict distinctly different present-day stellar masses.}
	\label{fig:host_met_distribution}
\end{figure}

In Fig.~\ref{fig:host_met_distribution} we present the mass weighted distribution of stellar metallicities, normalised such that the integral is unity. Here we show the distribution for halo G42, with the other systems showing similar trends. Here we again present the median lines of the MCMC chains, split into early, mid and late formations (see end of Section~\ref{section:host_stellar_mass}). Additionally shown for reference is a `weak' and `strong' feedback scenario. These use the stellar feedback parameters of $f_{\max} = 10$, $A = 0.5$ (strong feedback) and $f_{\max} = 0.5$, $A = 0$ (weak feedback), all other parameters are fixed to their fiducial values (see Table~\ref{Table:emulation_parameters}). These choices of stellar feedback parameters lead to very different present-day stellar masses, with $M_{\rm{\ast}} = 5.9 \times 10^{8}$ and $1.4 \times 10^{11}$ M$_{\rm{\odot}}$ for the strong and weak scenarios, respectively, whereas the different formation scenarios are constrained to have $M_{\rm{\ast}} \sim 10^{10}$ M$_{\rm{\odot}}$. As such, these are not realistic choices, but do show the possible effects of changes to stellar feedback, as well as what can be sampled using the emulator.

For the strong and weak feedback choices there are clear differences in the metallicity distributions, with strong feedback suppressing star formation, leading to a lower total stellar mass that overall has less enrichment and a lower metallicity.  The opposite is true for weak feedback. When considering the selection based on stellar formation time, with a fixed present-day stellar mass, the differences in the distributions are minimal. In particular, any systematic changes are well within the scatter (grey band). This suggests that the dominant factor in setting the metallicity, both averaged and the overall distribution, is the total number of stars that have formed, with the details of how these are formed being of secondary importance.

In this analysis we have only studied the total metallicity distribution. Notably, not splitting stellar particles into the different components of the galaxy (i.e. bulge, disk, halo, etc.). It is therefore likely that strong signals could be found with a more detailed analysis, which we leave for future work.

\subsection{Satellite stellar mass function}

\begin{figure*}
	\centering
	\includegraphics[width=\textwidth]{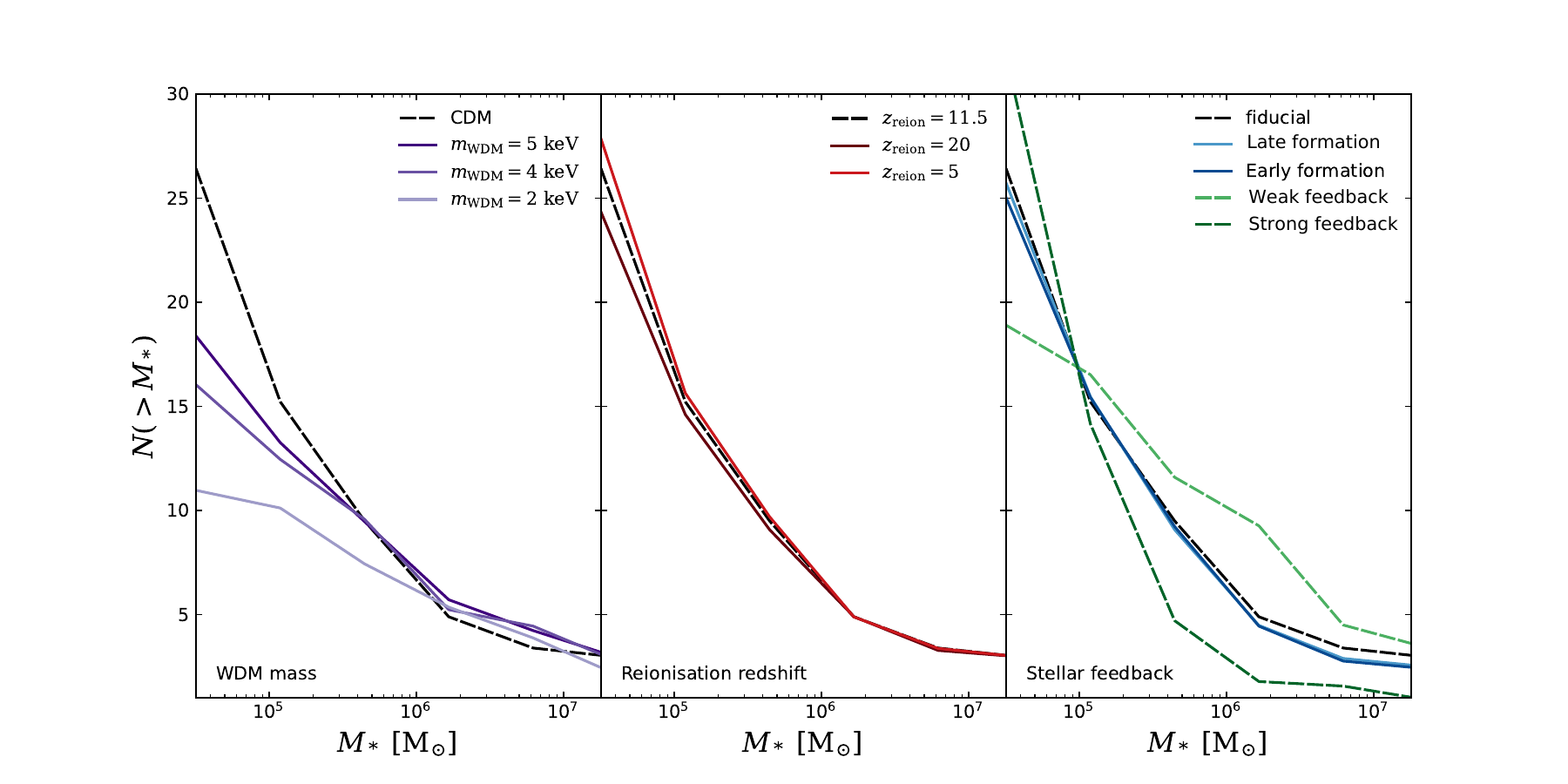}
	\caption{The cumulative satellite stellar mass function per halo, averaged over the three sampled systems. Each panel varies one (set of) parameters at a time, with all other parameters fixed to their fiducial values. Left panel changes the assumed WDM mass, $m_{\rm{WDM}}$, the middle the reionisation redshift and the right panel the stellar feedback. The stellar feedback is split into early, mid and late stellar formation (see Fig.~\ref{fig:host_stellar_mcm_form_time}) that all have comparable $z=0$ host stellar masses, and plotted for comparison is a strong and weak feedback scenario. Throughout, the fiducial CDM result is plotted as a dot-dashed black line.}
	\label{fig:sat_lum_func}
\end{figure*}

While it is important to understand the role of the different feedback parameters in changing properties of the host galaxy, these are generally not sensitive changes in the WDM mass, making them poor probes to constrain the WDM particle mass (for example, see Fig.~\ref{fig:host_stellar_mass_param_dep}), or other similar small scale deviations from $\Lambda$CDM. It is expected, and indeed we find, that the properties of the satellite population to be more sensitive to deviations from CDM \citep[e.g.][]{Lovell_14, Stafford_20,Forouhar_22}. While it is possible to study and emulate many different properties of the satellites, here we focus on the satellite stellar mass function. Where the host stellar mass is predominantly set by the three stellar feedback parameters, the number of luminous satellites is sensitive to both the stellar feedback, the reionisation redshift and the WDM mass. It is therefore important to understand and quantify how the freedom in the baryonic parameters impact the constraints on WDM. Here we present the effect of changing the different parameters in isolation to help develop an intuition for the different roles of feedback, reionisation and the suppression of the initial density field on the luminous satellite population. In the future we plan to study joint changes to these parameters.

We define the satellite stellar mass function by selecting all subhaloes within $300$kpc of the host and use the stellar mass identified by \texttt{SUBFIND}. We then present the cumulative stellar mass function using 10 logarithmically-spaced bins in the range $M_{*} = 2.23 \times 10^4$ -- $10^{9.5}$ M$_{\rm{\odot}}$ and excluding the host.

In Fig.~\ref{fig:sat_lum_func} we present the cumulative stellar mass function averaged over all three systems. The left panel shows the effect of varying the WDM mass and the middle panel shows the effect of varying the redshift of reionisation. In both cases all the other parameters are held fixed. The right panel explores the effects of varying the three stellar feedback parameters for the CDM case and using the fiducial reionisation redshift. Here we show the averaged stellar mass function when fitting to the host stellar mass (i.e. Fig.~\ref{fig:host_stellar_mcm}) with one sigma uncertainty, as well as the average when split into early and late forming hosts (as defined previously). Additionally plotted for reference are a `strong' and `weak' feedback scenarios (see section for specific values). These choices of parameters predict significantly different stellar masses for the host, that are not consistent with the SMHM relation inferred from abundance matching.

Focusing initially on the effect of changes to the assumed WDM mass, $m_{\rm{WDM}}$, (left panel of Fig.~\ref{fig:sat_lum_func}) we see that a smaller particle mass results in a suppression of number of observed satellites at low stellar masses, with the mass scale that these differences occur being sensitive to $m_{\rm{WDM}}$. This suppression is expected, as WDM leads to a suppression in the initial power spectrum (see Eqn.~\ref{eqn:wdm_suppression}) leading to a suppression in the number of DM (sub)haloes and in turn a suppression in the luminous satellites. In general, WDM can have a measurable effect across a wide range of mass scales, assuming a small enough WDM particle mass. However, with current conservative constraints suggesting $m_{\rm{WDM}} \gtrsim 2$ keV \citep[e.g.][]{Newton_21} the effects of WDM are only significant in the mass range $M_{\rm{\ast}} \lesssim \times 10^6 \rm{M}_{\odot}$.

We now focus on the role of reionisation in changing the stellar mass function (middle panel Fig.~\ref{fig:sat_lum_func}). The first important thing to note is that reionisation only affects the smallest galaxies, with the mass range being similar to that of WDM ($M_{\rm{star}} \lesssim 10^5 \rm{M}_{\odot}$). Massive haloes offer a large enough gravitational potential to retain their gas after reionisation, while smaller haloes lose most of their gas once heated \citep[e.g.][]{Benitez-Llambay_2020}. The exact mass scale is debated but is roughly $M_{200c} \sim 10^7$ M$_{\rm{\odot}}$, corresponding to a stellar mass of $M_{*} \sim 10^5$ M$_{\rm{\odot}}$.\footnote{In general, stellar mass will depend on the assumed stellar mass--halo mass relation for dwarf haloes, which itself will depend on the assumed feedback efficiencies.} The observed trend is that a later reionisation leads to the formation of more dwarf galaxies at low masses ($M_{*} \lesssim 10^5 \rm{M}_{\odot}$), and vice versa. This dependence is readily explained by assuming that before reionisation these systems are actively star forming and that reionisation directly quenches them. Therefore, if reionisation happens later these systems have more time to form stars prior to reionisation, resulting in larger stellar masses and an increased number of galaxies at these mass scales. The magnitude of the effect over the sampled redshift and mass ranges is relatively small, only a few per system on the total number counts.

In the right panel of Fig.~\ref{fig:sat_lum_func} we explore how the satellite stellar mass function is affected by variations to stellar feedback. We consider choices of parameters that give a consistent host stellar mass, shown as solid lines, split into late and early formation scenarios, as described in section \ref{section:host_stellar_mass}. Finally, for comparison we also plot a strong and weak feedback model (see Section \ref{section:met_dist} for the specific combination of parameters).

Focusing initially on the lines where the host stellar masses are fixed (solid line), we see that there is little dependence on formation time, with any deviations well within the intrinsic scatter and uncertainty on the emulator. Additionally, we find that there is almost no strong correlation with the formation time of the stellar component of the host. If we now ignore the host stellar mass and just consider the strong and weak feedback scenarios (dashed lines) as examples of what is possible then we see that stellar feedback is able to significantly change the stellar masses of the satellites. And in general effects the whole stellar mass range, where it is not possible to make changes to isolated mass scales. At high masses ($M_{\rm{\ast}} \gtrsim 10^5$ M$_{\odot}$) the effect is as expected, where stronger feedback leads to a reduction in the number of satellites, and vice versa. However, at small masses ($M_{\rm{\ast}} \lesssim 10^5$ M$_{\odot}$) we see this trend reverse so that strong feedback leads to an increase in the total number of luminous satellites. This behaviour appears to be driven by interactions with the host; in a strong feedback scenario the host system forms comparatively fewer stars, hence reducing the tidal stripping of satellites, leading to an increase in the number of satellites with small stellar mass. The reverse of this applies to the weak feedback scenario, where the host forms considerably more stars, increasing the disruptive effects from the host, such as tidal stripping. We have verified this hypothesis by also studying the satellite DM mass function that shows a decrease in the number of subhaloes over all mass scales in the strong feedback scenario, and an increase in the weak feedback scenario, relative to the fiducial case. Clearly showing that the overall amount of substructure is affected, not just how those haloes are populated with luminous galaxies. However, to conclusively show that it is the effects of interactions with the host would involve matching (sub)haloes across the different runs and studying their evolution after accretion, which is beyond the scope of this work.

It is clear that all three processes play a role in setting the observed satellite populations. It is therefore important to consider potential degeneracies between the baryonic processes modelled here and changes to the nature of DM. WDM and reionisation, both affect the satellite stellar mass function over the same mass scales and the form of the effect is similar. The key difference is that WDM can only suppress the number of satellites, while changes to the reionisation redshift can either relatively enhance or suppress satellite growth. However, the magnitude of their effects are significantly different. There is therefore only a mild degeneracy between the reionisation redshift and WDM. Stellar feedback is able to have the same magnitude of an effect as WDM, though the form of the change is distinct, with changes to stellar feedback tending to affect the whole stellar mass function while the effects of WDM free-streaming tend to only be important below a mass scale that is determined by the WDM particle mass. While the total number of luminous satellites above a given mass threshold is degenerate between the two processes, this degeneracy can be broken by studying the full stellar mass function where the effects of WDM and stellar feedback are distinct. Additionally, if the host stellar mass is also constrained, there is significantly less freedom in changing the luminosity function.

\section{Summary} \label{section:summary}

In this work we have presented a new suite of high resolution cosmological zoom-in simulations of Milky Way-mass haloes where key model parameters are systematically and simultaneously varied. Three haloes from the existing ARTEMIS simulations have been resimulated many times, with different assumptions about the WDM mass and the baryonic physics parameters (Fig.~\ref{fig:visualisation}). In total, six parameters are simultaneously and systematically varied: the WDM mass, the reionisation redshift, the star formation gas density threshold and three parameters associated with stellar feedback. From these simulations, emulators have been built (Section~\ref{section:emulator}, Fig.~\ref{fig:emulator_summary}) for a wide range of statistics from the simulations (currently there are approximately $250$ unique summary statistics trained), such as the host stellar mass or the number of satellites, to be predicted as a function of the $6$ varied parameters, $\vec{\theta} = (m_{\rm{DM}}, A, f_{\rm{max}}, \rho_{\rm{H,0}}, n^*_{\rm{H,0}}, z_{\rm{reion}})$. In this first paper we have primarily focused on emulating a range of summary statistics, however the new simulation suite is well suited for developing more advanced machine learning techniques, such as deep learning and likelihood free inference.

The emulators allow for both the cosmological and baryonic parameters to be simultaneously varied. The significant increase in computational speed offered by the emulator compared to directly running the simulations, roughly a factor of $10^{11}$, allows for a full exploration of the $6$-dimensional space, as opposed to being fixed to pre-calibrated values as is typical in the literature.

In this paper we focused on presenting the simulations and emulators, along with demonstrating some of the possible applications of this new approach and exploring the role of feedback and cosmology on a handful of common statistics. The analysis and results can be summarised as follows:

(i) We study how the stellar mass of the host (i.e. the Milky Way analogue) varies as a function of the emulated parameters (Fig.~\ref{fig:host_stellar_mass_param_dep}). It is found that the stellar mass is most sensitive to the three stellar feedback parameters, with possible changes of an order of magnitude from the fiducial case, while the assumed reionisation redshift and warm darker matter mass have a negligible effect.

(ii) We additionally perform an MCMC analysis, fitting the stellar mass of the host to the stellar mass--halo mass relation from \citealt{Behroozi_19}. Strong degeneracies in the stellar feedback parameters are identified (Fig.~\ref{fig:host_stellar_mcm}). We further explore the physical origin of these degeneracies by studying the redshift evolution of the progenitor. Here it is found that the degeneracy in the feedback parameters corresponds to a freedom in the formation time of the stellar component (Fig.\ref{fig:host_stellar_mcm}). We additionally split the MCMC chains into three formation scenarios (early, mid and late), corresponding to systematic changes to the input parameters.

(iii) Additional statistics beyond the stellar mass are explored, including the mean metallicity, the half mass radius, $r_{1/2}$, the in situ fractions and the stellar morphology (Fig.~\ref{fig:host_additional_stat}). It is found that present-day metallicity and stellar morphology are broadly insensitive to the stellar formation time, while the host size (i.e. stellar half mass radius) and in situ fractions demonstrate clear systematic trends with formation time. A late formation scenario corresponds to an increased stellar half-mass radius and an increased in situ fraction.

(v) Finally, we explore the isolated effect of changes in the stellar feedback, reionisation redshift and WDM mass on the satellite stellar mass function (Fig.~\ref{fig:sat_lum_func}). Here it is found that changes to the reionisation redshift (over the range $z_{\rm{reion}} = 5$--$20$) has a minimal effect on the number of luminous satellites above $M_{*} \sim 10^4$ M$_{\rm{\odot}}$, with deviations $\sim 2$ per system. Variations in the WDM mass lead to a suppression in the number of satellites at small stellar masses, $M_{*} \lesssim 10^{6}$M$_{\rm{\odot}}$ compared to CDM. Variations in stellar feedback parameters are able to suppress or enhance the total number of satellites, with changes of a similar magnitude to that of WDM, but are not isolated to a particular mass scale. This analysis suggests that stellar feedback and WDM are not strongly degenerate with each other, and the satellite luminosity function of the Milky Way and similar systems can be a powerful probe of \textit{both} galaxy formation and cosmology. We plan to explore this further in future work.

In summary, the emulators allow for fast ($\sim 1$ms) predictions for a diverse range of statistics as a function of both cosmological and baryonic (feedback) parameters. The significant increase in computation speed (a factor of $\sim 10^{10}$) alleviated one of the key limitations of standard cosmological hydrodynamic simulations; the high computational expense. This fundamentally changes the type of analysis that can be performed. In particular, it is now possible to fully explore the available parameter space, and perform Bayesian inference analysis using Markov-chain Monte Carlo analysis, and similar methods. While this significantly increases the predictive power of these simulations, allowing for their model (subgrid) parameters to be marginalised, it also allows for a deeper understanding of the link between the models used and the resulting galaxy properties. We hope that these emulators will become an invaluable tool to further understand the role of baryonic process and cosmology in the formation and evolution of galaxies.

\section*{Acknowledgements}

The authors thank the referee for an insightful and constructive report that helped improve the clarity and quality of the final manuscript. This project has received funding from the European Research Council (ERC) under the European Union’s Horizon 2020 research and innovation programme (grant agreement No 769130). S.T.B. and A.F. are supported by a UKRI Future Leaders Fellowship (grant no MR/T042362/1). K.A.O acknowledges support by the Royal Society through Dorothy Hodgkin Fellowship DHF/R1/231105. A.H.R.~is supported by a Research Fellowship from the Royal Commission for the Exhibition of 1851. This work used the DiRAC@Durham facility managed by the Institute for Computational Cosmology on behalf of the STFC DiRAC HPC Facility (www.dirac.ac.uk). The equipment was funded by BEIS capital funding via STFC capital grants ST/K00042X/1, ST/P002293/1, ST/R002371/1 and ST/S002502/1, Durham University and STFC operations grant ST/R000832/1. DiRAC is part of the National e-Infrastructure.

\section*{Data Availability}

All simulations presented here, along with the pre-trained emulators, are available upon a reasonable request to the corresponding author.



\bibliographystyle{mnras}
\bibliography{References} 




\appendix

\section{Accuracy and stochasticity test} \label{section:accuracy_appendix}

\begin{figure}
	\centering
	\includegraphics[width=0.485\textwidth]{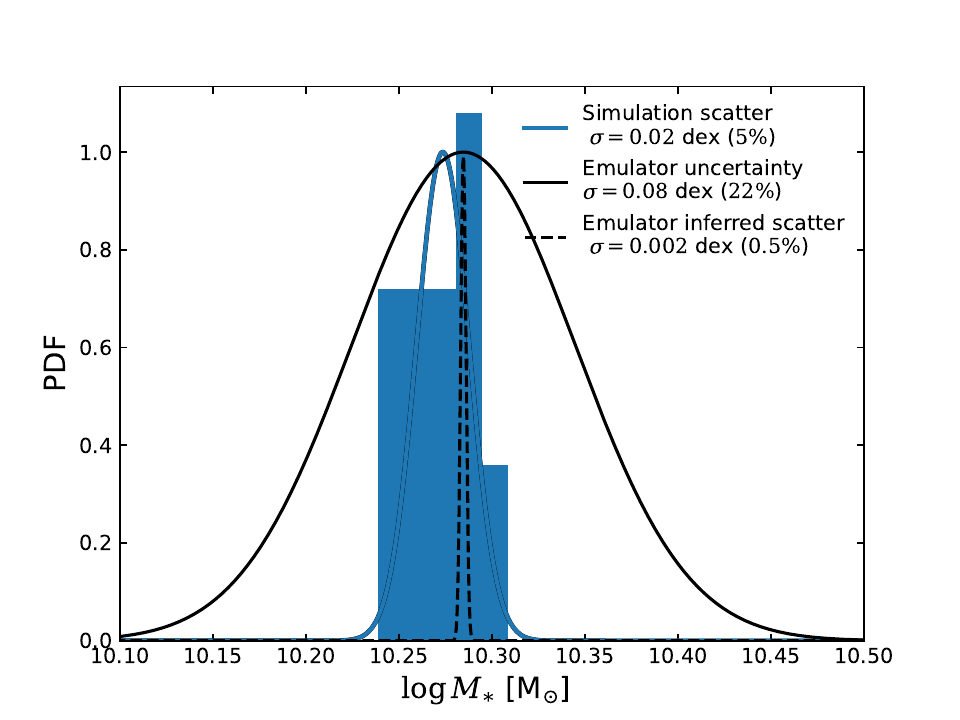}
	\caption{The distribution of $z=0$ stellar masses for G42 using the fiducial combination of parameters but changing the random seed used for the star formation and feedback models. The histogram shows the distribution of values, which closely follows a log-normal distribution. The blue line shows the Gaussian, with the same mean and standard deviation as the data. For comparison the emulator prediction for the mean is shown as the black solid line, and the inferred intrinsic scatter as the dashed black line. Note that the distributions have been normalised so that their maxima are unity for easier comparison.}
	\label{fig:stochasticity_test}
\end{figure}

\begin{figure}
	\centering
	\includegraphics[width=0.485\textwidth]{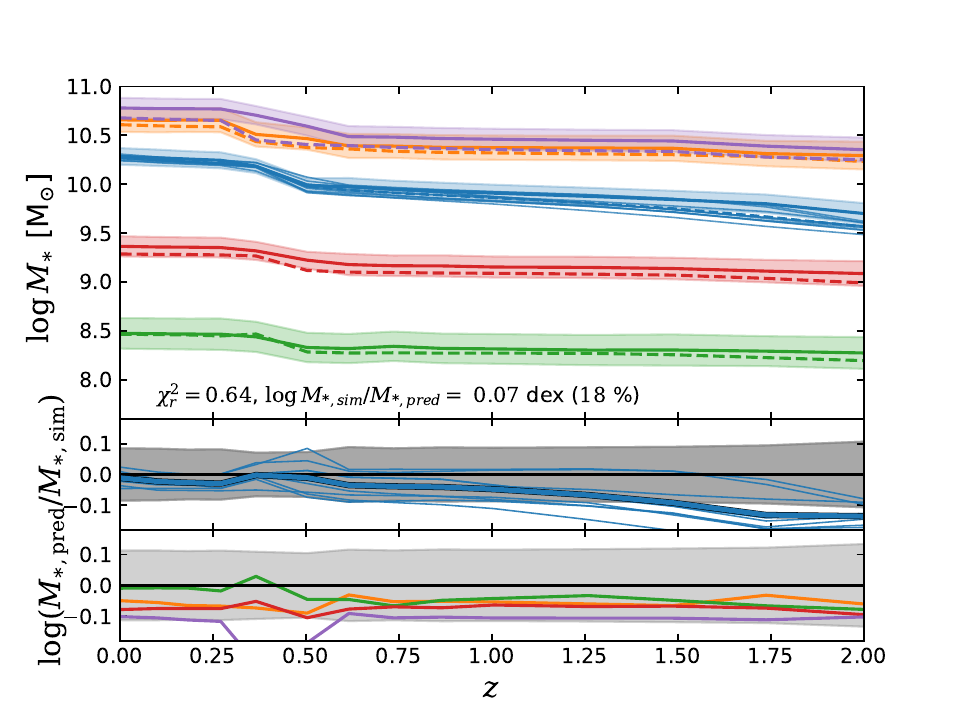}
	\caption{The redshift evolution of the host stellar mass, taken to be the mass within $30$kpc. The simulation outputs are shown as dashed lines, with the fiducial parameters shown in blue, including all reruns with the varied random seed, while the other colours show the outputs for the four random combinations of parameters used to evaluate the accuracy of the emulator. Additionally plotted is the emulator prediction as the solid lines, with the shaded region showing the corresponding uncertainty in the prediction.}
	\label{fig:accuracy_test}
\end{figure}

Modern cosmological hydrodynamic simulations make extensive use of probabilistic, Monte-Carlo based algorithms to model star formation and feedback processes. This inherent randomness, coupled with the chaotic orbits of individual particles, means that the simulations are not fully deterministic, with their outputs depending both on the choice of input parameters and the particular run. For statistics that average over a large number of individual systems, such as the stellar mass function, the impact of this inherent stochasticity is minimal. However, for individual systems the variation from different runs can be significant \citep[e.g.][]{Keller_19,Borrow_23,Davies_23}, depending on the quantity being compared between runs, the nature of the subgrid modelling, the resolution, and the formation history of a given system.

To explore this inherent stochasticity in our simulations and the effect it has on both their predictive power and the ability to train emulators from individual runs, we have rerun G42 $10$ times with the fiducial choice of parameters (see Table.~\ref{Table:emulation_parameters} for values), each time changing the random seed used. We present the results for this in Fig.~\ref{fig:stochasticity_test} for the stellar mass of the host at $z=0$. The histogram of the stellar masses is shown in blue, where it can be well fit by a log normal distribution, with the best fit Gaussian shown in the solid blue line. The standard deviation is $\sigma = 0.02$ dex ($\approx 5 \%$), showing that the present-day stellar mass is robustly determined in these simulations. This is notably smaller than found in other works \citep[e.g.][]{Borrow_23}, and is likely due to these simulations being of significantly higher resolution (a factor of $\approx 60$ in particle mass).

Additionally plotted in Fig.~\ref{fig:stochasticity_test} is the emulator's prediction for the stellar mass at the fiducial choice of parameters (not used to train the emulator). The emulator is constructed such that the data is assumed to have some intrinsic scatter. The prediction for the emulator is then the mean of the distribution at the given choice of parameters, along with an error on predicting that mean. The solid black line shows the emulator prediction, assumed to be Gaussian in form, and accurately recovers the mean of the distribution. The uncertainty in making the prediction ($\approx 20 \%$) is significantly larger than the intrinsic scatter in the simulations ($\approx 5 \%$). As such, we are currently limited by the uncertainty in making the prediction, and not yet the intrinsic scatter in the simulations. The accuracy of the emulator could be improved by increasing the number of nodes used to sample the space, or alternatively using a similar number but using an alternative coordinate system for the input parameters so that we do not sample as extreme variations in the properties of the simulated galaxies.

As well as making a prediction for the mean with a corresponding uncertainty, the emulator aims to infer the intrinsic scatter in the data. This prediction is shown in the dashed black line ($\sigma \approx 0.5 \%$), which under predicts the true value. This is likely due to the uncertainty on making the prediction being significantly larger than the intrinsic scatter. Additionally, this has no impact on any analysis using the emulator, as the emulator is the dominant uncertainty and will therefore dominate any likelihood analysis.

To further study the accuracy of the emulator we compare the predictions for the stellar mass as a function of redshift. This is shown in the top panel of Fig.~\ref{fig:accuracy_test}, where we present the simulation results for both the fiducial combination of parameters with all ten realisations alongside the four hold out tests that represent random combinations of parameters within the emulation range. The simulation results are shown as dashed lines, with the colours showing the different choices of parameters (see legend). The prediction for the emulator, along with the uncertainty, is shown in the solid lines. The two bottom panels show the ratio between the predicting and the simulations, split into the multiple realisations of the fiducial run and the four hold out tests.

In general the agreement between the simulations and the emulator is good. The absolute error from the emulator and hold out tests is $\approx 0.1$dex, and importantly any deviations are within the predicted uncertainty. Over the majority of the redshift range sampled deviations are within $1 \sigma$, with a few  deviations by approximately 2$\sigma$. To quantify the agreement we calculate the reduced $\chi^2$ which is found to be $\chi^2_r = 0.68$, showing an excellent fit to the data. Generally, it is expected that $\chi^2_r \approx 1$ for a good fit to the data, with $\chi^2_r < 1$ normally suggesting an overfit to the data. However, here we are comparing choices of parameters not used to develop the model, and therefore are independent. Therefore, the good agreement between the simulation and emulator suggests overfitting is not an issue in this case. Instead, it appears that the predicted uncertainties are larger than the true values. Therefore, using the uncertainties from the emulator in any statistics analysis places a conservative constraint on the predictive power of the model and emulator, and crucially prevents over interpreting the results of the emulator due to under-predicting the uncertainty.

In Fig.~\ref{fig:accuracy_test} (middle panel) it is also observed that the intrinsic scatter in the simulations is correlated, where realisations that have formed more stars by today also tended to have higher stellar masses at early times. However, the fractional scatter tends to decrease with time, such that there is a much larger scatter at $z \sim 2$ than today. This suggested that, while these systems are affected by the butterfly effect, they tend to become self regulating, leading to a similar present day stellar mass (at least within $\approx 5 \%$). Currently, these correlated errors are not taken into account when training the emulator. However, as discussed in the previous paragraph we are currently not limited by the intrinsic scatter of the simulations, so this should not have a significant effect on the accuracy of the emulator.

In conclusion the emulator offers an accurate prediction for the outputs of the simulations, including the intrinsic variation to the simulations. For the host stellar mass it is found that the intrinsic scatter between various simulation runs is $\approx 5 \%$, the absolute error on the emulator is $\approx 0.1$dex and provides reliable uncertainties. While we have only shown this analysis for the host stellar mass, we find that same conclusions for a wide range of other properties, such as the host metallicity, size, and even satellite counts. However, the exact values for the intrinsic simulation scatter and absolute errors on the emulator vary depending on the given statistic, with the deviations always within the predicted errors.


\bsp    
\label{lastpage}
\end{document}